\documentclass[
reprint,
twocolumn,
superscriptaddress,
amsmath,amssymb,
aps,
prx,
floatfix,
]{revtex4-2}

\usepackage{graphicx}
\usepackage{dcolumn}
\usepackage{bm}
\usepackage{braket}
\usepackage{siunitx}
\usepackage[dvipsnames]{xcolor}
\usepackage[colorlinks=true, allcolors=blue]{hyperref}

\usepackage{cleveref}
\crefname{equation}{Eq.}{Eqs.}
\Crefname{equation}{Equation}{Equations}
\crefname{figure}{Fig.}{Figs.}
\Crefname{figure}{Figure}{Figures}
\crefname{section}{Sec.}{Sects.}
\Crefname{section}{Section}{Sections}
\crefname{table}{Table}{Tables}
\crefname{appendix}{Appendix}{Apps.}
\Crefname{appendix}{Appendix}{Apps.}

\newcommand{\bd}[1]{{\color{Black}{#1}}}  % For submission only

\newcommand{\rr}[1]{{\color{Black}{#1}}}  % Highlight for referees round 1.
\newcommand{\rrr}[1]{{\color{Black}{#1}}}  % Highlight for referees round 2.

\begin{document}

\title{Probing Excited-State Dynamics of Transmon Ionization}
\author{Zihao~Wang}\thanks{Present address: AWS Center for Quantum Computing,
Pasadena, CA, USA}
\affiliation{Department of Physics and Astronomy, University of Rochester, Rochester, New York 14627}
    \affiliation{University of Rochester Center for Coherence and Quantum Science, Rochester, New York 14627, USA}
    
\author{Benjamin~D'Anjou}
\affiliation{Institut Quantique and D\'epartement de Physique, Universit\'e de Sherbrooke, Sherbrooke J1K 2R1 Québec, Canada}

\author{Philippe~Gigon}
\affiliation{Institut Quantique and D\'epartement de Physique, Universit\'e de Sherbrooke, Sherbrooke J1K 2R1 Québec, Canada}

\author{Alexandre~Blais}
\affiliation{Institut Quantique and D\'epartement de Physique, Universit\'e de Sherbrooke, Sherbrooke J1K 2R1 Québec, Canada}
\affiliation{CIFAR, Toronto, M5G 1M1 Ontario, Canada}

\author{Machiel~S.~Blok}
\email[]{machielblok@rochester.edu}
\affiliation{Department of Physics and Astronomy, University of Rochester, Rochester, New York 14627}
    \affiliation{University of Rochester Center for Coherence and Quantum Science, Rochester, New York 14627, USA}
\date{\today}

\begin{abstract}
The fidelity and quantum nondemolition character of the dispersive readout in circuit QED are limited by unwanted transitions to highly excited states at specific photon numbers in the readout resonator. This observation can be explained by multiphoton resonances between computational states and highly excited states in strongly driven nonlinear systems, analogous to multiphoton ionization in atoms and molecules. In this work, we utilize the multilevel nature of high-$E_J/E_C$ transmons to probe the excited-state dynamics induced by strong drives during readout. With up to 10 resolvable states, we quantify the critical photon number of ionization, the resulting state after ionization, and the fraction of the population transferred to highly excited states. Moreover, using pulse-shaping to control the photon number in the readout resonator in the high-power regime, we tune the adiabaticity of the transition and verify that transmon ionization is a Landau-Zener-type transition. \rr{We further extend these methods to a typical transmon with $E_J/E_C \approx 55$ and probe the offset-charge dependence of ionization dynamics in a timed-resolved manner.}
Our experimental results agree well with the theoretical prediction from a semiclassical driven transmon model and may guide future exploration of strongly driven nonlinear oscillators. 
\end{abstract}

\maketitle

%%%%%%%%%%%%%%%%%%%%%%%%%%%%%%%%%%%%%%%%
\section{\label{sec: intro}Introduction}
% This paragraph answers why we do this.
The ability to perform fast, high-fidelity, quantum nondemolition (QND) measurements is essential for quantum error correction and, more generally, for any quantum circuit that requires mid-circuit measurement. The standard method for qubit measurement in superconducting circuits is dispersive readout~\cite{blais2021, blais2004}. In this approach, a superconducting qubit, e.g., transmon~\cite{koch2007} or fluxonium~\cite{manucharyan2009}, weakly coupled to a far-detuned resonator, induces a state-dependent frequency shift to the resonator. A qubit measurement is performed by exciting the resonator with a readout tone, such that the field in the resonator entangles with the qubit, resulting in a projection of the qubit state as the field is detected~\cite{krantz2019}. In principle, this process is QND, and the signal-to-noise ratio within a given time can be improved by increasing the amplitude of the resonator field. Recent experiments have achieved over $99\%$ assignment fidelity on transmons with readout times equal to or less than \qty{100}{\ns}~\cite{walter2017, sunada2022, swiadek2024, sunada2024, hazra2025, spring2025}. Despite this progress, readout errors continue to be a major bottleneck in achieving fault-tolerant quantum computation~\cite{acharya2023, acharya2025}. 

% This paragraph lists what has been done.
An important limitation to dispersive readout in superconducting circuits is that strong drives can excite the device outside its computational two-level subspace~\cite{jeffrey2014, sank2016, walter2017, Minev2019, Verney2019, khezri2023, hazra2025, kurilovich2025, connolly2025, dai2026, fechant2025, xia2025} in a process that has been referred to as measurement-induced state transition (MIST)~\cite{sank2016} and transmon ionization~\cite{shillito2022, Verney2019, Mathieu_thesis}. Multiphoton qubit-drive resonances have been identified as a source of these transitions~\cite{sank2016}, and theoretical frameworks capable of predicting their occurrence have been developed~\cite{sank2016, shillito2022, cohen2023, Xiao2023, khezri2023, dumas2024, pan2025}. These tools have also been applied to the fluxonium~\cite{nesterov2024b, singh2025, bista2025} and to develop methods for mitigating ionization~\cite{bengtsson2024, kurilovich2025, chapple2025b, chapple2025a, li2025b, mori2025a}. This phenomenon bears some resemblance to multiphoton ionization in atoms and molecules
%, where a strong laser or microwave field promotes the electron from a bound state into the continuum
~\cite{Mainfray1991, agostini1968, mainfray1980, martin1976, deng1984, deng1985, breuer1989}. In both cases, the system can be driven into a highly excited state with delocalized wavefunctions and energies above the confining potential.

% This paragraph emphasizes what we do differently.
A comprehensive understanding of these multiphoton processes is a key step in developing strategies to avoid unwanted transitions in dispersive readout. While experimental results are consistent with theoretical predictions for the critical photon number of transmon ionization, other features---such as the final state reached and the occurrence of Landau-Zener dynamics---remain unverified. It is challenging to observe these phenomena since the control and measurement of typical transmons are often limited to the 4 lowest states, excluding the highly excited states. In this work, we study these unexplored features of transmon ionization by directly measuring its excited-state dynamics using high-$E_J/E_C$ transmons that enable high-fidelity control and readout of 10 energy eigenstates~\cite{wang2025, champion2025}. In the regime of negative transmon-resonator detuning, we demonstrate that the transmon ionization is indeed a pairwise transition between a qubit state and a highly excited state. We identify which states are populated, the critical photon number at which ionization happens, and the amount of population transfer during ionization. We find that both semiclassical dynamics simulations and Floquet analysis are in excellent agreement with our experimental results, confirming that the semiclassical driven transmon model can successfully capture the main features of ionization dynamics. Moreover, using pulse-shaping techniques to control the photon number in the resonator, we tune the system through the multiphoton resonance condition at variable speed and verify that the transmon ionization is a Landau-Zener-type transition where more population is ionized during an adiabatic process. \rr{Finally, we extend these experimental techniques to a typical transmon with $E_J/E_C \approx 55$. In this device, the critical photon numbers $n_{r,\rm crit}$ fluctuate over time and show a clear dependence on the offset charge, which is also well captured in our simulations.}

%%%%%%%%%%%%%%%%%%%%%%%%%%%%%%%%%%%%%%%%
\section{\label{sec: concepts}Ionization of high-$E_J/E_C$ transmons}

\begin{figure}[tp]
\includegraphics{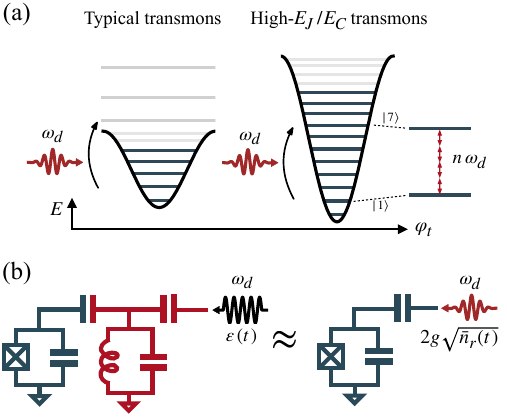}
\caption{
\label{fig: 1_concepts} 
Transmon ionization concepts. (a) The potentials and eigenstates of transmons. \rr{Here, we compare two transmons with $\omega_{01}/2\pi=\qty{5}{\GHz}$ and $E_J/E_C=55$ (left) or $E_J/E_C=275$ (right). These parameters are chosen for illustrative purposes. The actual parameters of our experimental devices can be found in \cref{sec: device_params}.}
The final state of ionization for a typical transmon is often close to (or even above) the top of its potential. High-$E_J/E_C$ transmons have a deeper potential and confine more energy levels, which makes the highly excited states accessible during the transmon ionization. The level diagram depicts a multiphoton resonance. In this example, the energy levels of states $\ket{1}$ and $\ket{7}$ are ac-Stark shifted by the drive to reach the resonance condition $\tilde{\omega}_7-\tilde{\omega}_1=n\omega_d$ at a certain drive power, with $\omega_d$ the drive frequency and $n$ the number of absorbed photons. Typically, $n > 1$. (b) Circuit diagrams. When the transmon is in one of its eigenstates, a readout pulse with frequency $\omega_d$ and amplitude $\varepsilon(t)$ creates a coherent state in the resonator. This coherent state can be effectively modeled as a classical drive applied directly to the transmon, which can induce transitions between transmon states. This driven transmon model is used for the numerical simulations in this work; see \cref{eq:driven_harmonic_transmon_hamiltonian}.
}
\end{figure}

The mechanism of transmon ionization can be understood as a multiphoton resonance in a driven transmon. The drive induces an ac-Stark shift to each transmon eigenstate, resulting in a resonance when the energy difference between two shifted transmon states equals an integer number of the drive photon energy~\cite{sank2016, shillito2022, cohen2023, Xiao2023, khezri2023, dumas2024}. As a result, a driven transmon can transition from its computational subspace to a highly excited state at a specific drive amplitude. In principle, these resonances can occur between many pairs of transmon states and for a variety of transmon parameters. In practice, however, typical transmons that have relatively shallow potentials are often excited to a state close to the top of the potential, as shown in \cref{fig: 1_concepts}(a), and such a highly excited state is harder to address experimentally due to charge noise. Transmon ionization to highly excited states has thus far only been observed indirectly as a leakage out of the qubit subspace. In contrast, the high-$E_J/E_C$ transmons in our experiments have deeper potentials and confine more energy levels; see the right-hand side of \cref{fig: 1_concepts}(a). As a result, at least 10 transmon eigenstates are insensitive to charge noise and can be controlled and measured~\cite{wang2025, champion2025}. This enables us to directly probe excited-state dynamics of transmon ionization. On the \rr{right-hand-side} of \cref{fig: 1_concepts}(a), we \rr{also} show an example level diagram where the transmon levels are ac-Stark shifted, and $n$ photons in the drive are absorbed to cause a transition between $\ket{1}$ and $\ket{7}$, as will be the case in \cref{sec: experiments}.

The dynamics of the transmon-resonator system under an external drive is governed by the Hamiltonian ($\hbar=1$)~\cite{koch2007, blais2021}
\begin{equation}
\begin{split}
\hat{H}(t) & = 4 E_C (\hat{n}_t - n_g)^2 
- \sum_{m = 1}^M E_{Jm} \cos(m \hat{\varphi}_t) \\
& + \omega_r \hat{a}^\dagger \hat{a} - i g (\hat{n}_t - n_g) (\hat{a} - \hat{a}^\dagger) \\
& - i \varepsilon(t) \cos(\omega_dt)(\hat{a} - \hat{a}^\dagger).
\label{eq:transverse_coupling_hamiltonian}
\end{split}
\end{equation}
In this expression, $E_{Jm}$, $E_C$, $\hat{n}_t$, $\hat{\varphi}_t$ and $n_g$ are respectively the Josephson energies, the charging energy, the charge operator, the phase operator, and the offset charge of the transmon. Moreover, $\omega_r$ and $\hat{a}$ are the bare frequency and annihilation operator of the resonator, while $\varepsilon(t)$ and $\omega_d$ are the amplitude and frequency of the capacitive drive on the resonator, respectively. The resonator has a linewidth $\kappa$. In the dispersive regime, it inherits a transmon-state-dependent frequency shift $\chi_j$ as well as Kerr $K_{r, \ket{j}}$ and higher-order nonlinearities for any nonzero coupling strength $g$~\cite{blais2021}. \Cref{eq:transverse_coupling_hamiltonian} includes higher harmonics of the Josephson potential that provide a more accurate description of the transmon spectrum~\cite{willsch2024, wang2025}. Two high-$E_J/E_C$ transmons, $Q_A$ with $E_{J1}/E_C \approx 275$ and $Q_B$ with $E_{J1}/E_C \approx 235$, \rr{and a typical transmon $Q_C$ with $E_{J1}/E_C \approx 55$} are used in this work. For \rr{all} transmon-resonator pairs, the qubit frequency is lower than the resonator frequency; see \cref{sec: device_params} for the full set of parameters.

Because of this choice of qubit-resonator detuning and of the weak transmon-resonator coupling strength $g/2\pi \sim \qty{30}{\MHz}$, the dispersive shifts are small, and ionization occurs at large photon numbers in these devices~\cite{dumas2024}. As a result, the dimension of the full transmon-resonator Hilbert space required to model and simulate the experiment is prohibitively large. However, previous works have shown that the coherent state $\alpha(t)$ in the resonator generated by the readout tone approximately results in an effective classical drive acting on the transmon~\cite{cohen2023, lledo2023, khezri2023, dumas2024}; see \cref{fig: 1_concepts}(b). In that case, the effective semiclassical Hamiltonian for the transmon is
\begin{equation}
\begin{split}
\hat{H}_{\rm sc}(t) &= 4 E_C (\hat{n}_t - n_g)^2 
- \sum_{m = 1}^M E_{Jm} \cos(m \hat{\varphi}_t)  \\
&- 2g \sqrt{\bar n_r(t)} \sin[\omega_d t - \phi(t)] (\hat{n}_t - n_g), \label{eq:driven_harmonic_transmon_hamiltonian}
\end{split}
\end{equation}
where the resonator field is written as $\alpha(t) = \sqrt{\bar n_r(t)} e^{i \phi(t)}$ with $\bar n_r(t) = |\alpha(t)|^2$ being the average photon number. Transmon ionization occurs at specific values of $\bar n_r$, corresponding to a set of critical photon numbers \rr{$n_{r,\rm crit}$} for each transmon eigenstate $\ket{j}$. 
% This model neglects quantum fluctuations in the resonator and, in particular, measurement-induced dephasing. Nevertheless, as discussed below, it captures experimental observations after the results are averaged over offset charge.
% \bencomment{Get into why later. Another issue here is that I still need to connect this to the Kerr model used for calibration; it might be confusing.} \abc{I would not talk about this here. The reader is referred to an appendix for that, which seems good enough for the main text.}

%%%%%%%%%%%%%%%%%%%%%%%%%%%%%%%%%%%%%%%%
\section{\label{sec: experiments}Experimental identification of transmon ionization}

\begin{figure}[t]
\includegraphics{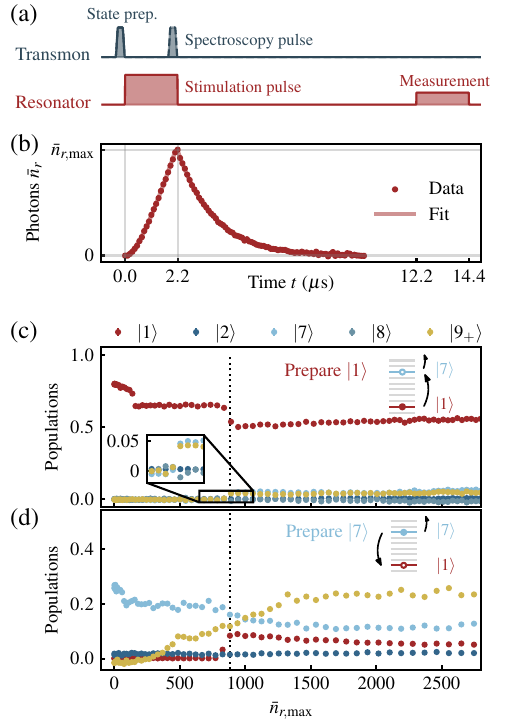}
\caption{
\label{fig: 2_experiments}
Transmon ionization experiments. (a) Pulse sequence for the ionization experiments. The transmon is prepared in one of its eigenstates $\ket{j}$ and then evolves under a \qty{2.2}{\us} stimulation drive on the resonator. After a \qty{10}{\us} ring-down, a weak multitone readout pulse is applied to measure the transmon populations. At low stimulation power, an optional spectroscopy pulse can be used to probe the mean photon number $\bar n_r$. (b) The measured $\bar n_r(t)$ for an experiment with $\bar n_{r, \rm{max}} \sim 150$. (c) Populations of the transmon under different stimulation amplitudes when it is prepared in $\ket{1}$. The vertical dashed line marks the critical photon number at which the $\ket{1} \leftrightarrow \ket{7}$ transition can happen. (d) Populations of the transmon under different stimulation amplitudes when initially prepared in $\ket{7}$. The ``deionization" shows the same critical photon number as the upward ionization. \rr{The error bars are smaller than the symbol size in this figure and all following figures.}
}
\end{figure}

We first show the excited-state populations for the ionization of $Q_A$. The sequence of our experiment is shown in \cref{fig: 2_experiments}(a). At the beginning of the sequence, the transmon is prepared in one of its eigenstates $\ket{j}$. Then, a \qty{2.2}{\us} square stimulation pulse is sent to the resonator. The frequency of this pulse is chosen to be on resonance with the dressed resonator frequency $\omega_d = \omega_{r, \ket{j}}$ at zero photon number. The stimulation is followed by a \qty{10}{\us} ($\sim 6.5/\kappa$) ring-down and then the end-sequence measurement. We calibrate the mean photon number $\bar n_r$ from the ac-Stark shift $ (\chi_{j+1} -\chi_j) \bar n_r$ using a \qty{40}{\ns} spectroscopy pulse on the transmon. By changing the timing of the spectroscopy pulse, the time-dependent $\bar n_r(t)$ can be measured; see \cref{fig: 2_experiments}(b). Because of the relatively small linewidth $\kappa$, the resonator does not reach a steady state during the stimulation pulse and requires a long ring-down time. In this experiment, we use this spectroscopy sequence to calibrate the conversion between stimulation amplitude and maximum mean photon number $\bar n_{r, \rm{max}}$ at low photon number, up to 400 photons, and extrapolate to higher photon numbers accounting for the induced Kerr nonlinearity. Details of the conversion and the effect of nonlinearity are discussed in \cref{sec: full_fig_2}. 

In \cref{fig: 2_experiments}(c), we show the resulting transmon populations at the end of the sequence for different maximum mean photon numbers $\bar n_{r, \rm{max}}$ when the transmon is initially prepared in $\ket{1}$. \rr{For clarity, here we only show the populations for the states that are of immediate interest for studying ionization, and the remaining populations are shown in \cref{sec: full_fig_2}. At $\bar n_{r, \rm{max}}=0$, the population in $\ket{1}$ is less than 1 because of the relaxation of the transmon during the \qty{12.2}{\us} sequence. This value is the reference population for no ionization.} We observe a series of distinct drops in the population of $\ket{1}$ at specific photon numbers. First, at $\bar n_{r, \rm{max}} \sim 170$, qubit $Q_A$ becomes resonant with a neighboring transmon, leading to an exchange of excitations between the two transmons; see \cref{sec: QAQBswap} for more details. More interestingly, a signature of ionization is visible at $\bar n_{r, \rm{max}} \sim 880$ where a population drop of $\ket{1}$ coincides with increased populations in several highly excited states. We attribute the fact that multiple excited states are populated to energy relaxation that occurs after ionization; \rr{see also \cref{sec: full_fig_2}}. We identify the highest resolvable excited state with a nonzero population as the final state, in this case, $\ket{7}$. Indeed, we do not observe population in $\ket{8}$ for this experiment. As will be shown in \cref{sec: comparisons}, state $\ket{7}$ is resonant with \rr{$\ket{18}$} below 880 photons, which could make the transferred population further ionize to higher excited states during \rr{the later stage of the experimental sequence}. These states are collectively classified as $\ket{9_+}$ and cannot be \rr{resolved} in our experiment.

Having experimentally characterized the transmon post-ionization state, we now proceed to further investigate the ionization process. Assuming the $\ket{1} \rightarrow \ket{7}$ transition occurs due to a multiphoton resonant process, the reverse ``deionization" process, $\ket{7} \rightarrow \ket{1}$, should also be observable at the same resonance condition, i.e., the same critical photon number. Consistent with this expectation, in \cref{fig: 2_experiments}(d), we prepare the transmon in $\ket{7}$ and indeed observe a population transfer from $\ket{7}$ to $\ket{1}$ at $\bar n_{r, \rm{max}} \sim 880$. The state $\ket{2}$ is not populated during this process, indicating a direct transition from $\ket{7}$ to $\ket{1}$. We also find a significant population increase in $\ket{9_+}$, which implies strong ionization to higher excited states.

%%%%%%%%%%%%%%%%%%%%%%%%%%%%%%%%%%%%%%%%
\section{\label{sec: comparisons}Comparisons with numerical simulations}

\begin{figure}[t]
\includegraphics{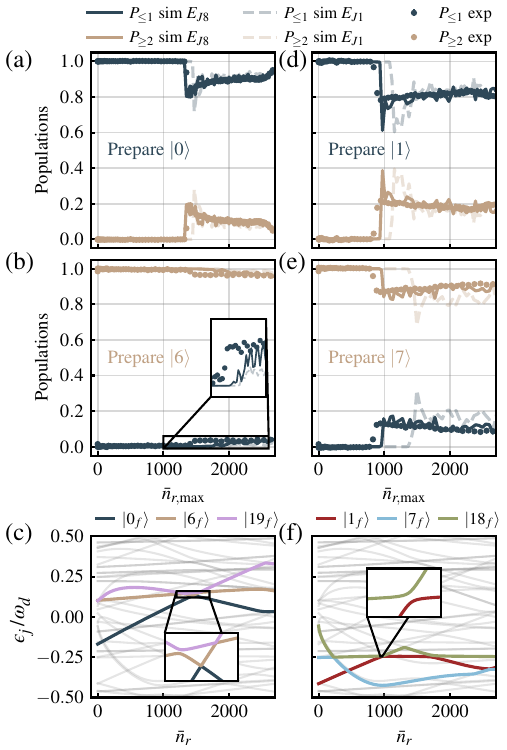}
\caption{
\label{fig: 3_comparisons} 
Comparisons between experiments and numerical simulations. (a-b) Transmon ionization associated with the $\ket{0} \leftrightarrow \ket{6}$ transition. Two different models, the conventional transmon model ($E_{J1}$, dashed lines) and the Josephson harmonics model ($E_{J8}$, solid lines), are used in simulations. We show the population of the qubit subspace $P_{\leq 1}$ and the populations of the higher excited states $P_{\geq 2}$. (c) Normalized Floquet quasienergies $\epsilon_j/\omega_d$ for each transmon branch when $\omega_d=\omega_{r, \ket{0}}$. The $\ket{0_f}$ branch has an avoided crossing with $\ket{6_f}$, which is also coupled to the higher-excited branch $\ket{19_f}$. (d-e) Similar to (a-b) but for the $\ket{1} \leftrightarrow \ket{7}$ transition. (f) Normalized Floquet quasienergies $\epsilon_j/\omega_d$ for each transmon branch when $\omega_d=\omega_{r, \ket{1}}$. The $\ket{1_f}$ branch has an avoided crossing with the $\ket{18_f}$ branch. The final state found in the dynamical simulations and the experiment is $\ket{7}$ instead of $\ket{18}$. This is due to a weak avoided crossing at lower photon number, at which most of the population in $\ket{18_f}$ is diabatically transferred to $\ket{7_f}$ during ramp-down. We take $n_g=0$ in (c) and (f).
}
\end{figure}

%% This paragraph tells the theoretical model and its assumptions, along with the reason to justify these assumptions.
We now compare the experimental results to theory. The coupled semiclassical equations for the transmon state $\ket{\psi(t)}$ and the resonator field $\alpha(t)$ in the rotating frame are
\begin{align}
    \ket{\dot{\psi}} &= -i \hat{H}_{\rm sc}(t) \ket{\psi}, \label{eq: schrodinger}\\
    \begin{split}\label{eq: alpha_dot}
        \dot{\alpha} &= -i(\omega_r - \omega_d) \alpha - \frac{\kappa}{2}\alpha - i\frac{\varepsilon(t)}{2} \\
        &\hspace{0.5cm} + g (\langle \hat{n}_t \rangle - n_g) e^{i\omega_d t}.
    \end{split}
\end{align}
The last term of \cref{eq: alpha_dot} describes the backaction of the transmon on the resonator field and captures all transmon-induced resonator nonlinearities within the semiclassical framework. Note that we performed the rotating-wave approximation on the resonator drive $\varepsilon(t)$. Our model does not include transmon relaxation or the weak interaction with the neighboring qubit described above, both of which are observed in the experiment. However, these processes are expected to be most relevant during the slow ring-down, after the multiphoton ionization that is of interest to us has already occurred. Indeed, the relaxation times of the most relevant states are longer or comparable to the duration ($\qty{12.2}{\us}$) of the pulse sequence\rr{; see \cref{sec: t1_correlation} for more details about the impact of $T_1$ on observed ionization result}. Moreover, we show in \cref{sec: QAQBswap} that for the high powers at which ionization occurs, population transfer to the neighboring qubit is negligible during the resonator ramp-up.

%% This paragraph tells what experimental data/results are shown.
To highlight the population transfer between the qubit subspace and higher excited states that occurs during the pulse sequence, we report the total population in the qubit subspace $P_{\leq 1}$ and the total population in the leakage subspace $P_{\geq 2}$. We plot these experimentally measured populations as a function of the maximum mean photon number $\bar n_{r, {\rm max}}$ in \cref{fig: 3_comparisons} (dots). We investigate two specific multiphoton resonances: the $\ket{0} \leftrightarrow \ket{6}$ resonance in \cref{fig: 3_comparisons}(a, b) and the $\ket{1} \leftrightarrow \ket{7}$ resonance in \cref{fig: 3_comparisons}(d, e).

%% This paragraph tells what theoretical data/results are shown.
The theoretical prediction is obtained by numerically solving the coupled semiclassical system in \cref{eq: schrodinger,eq: alpha_dot} \rr{using the {\sc JAX}-based {\sc Diffrax} library~\cite{kidger2021on}}. The simulations include the ring-down stage since the residual photons after stimulation can also induce ionization. This is especially true since the resonator field varies more slowly during ring-down, increasing the probability of ionization~\cite{shillito2022, dumas2024}. The simulation is performed using two distinct models of transmons: the conventional transmon model with a single Josephson harmonic ($E_{J1}$ model) and a model that includes 8 Josephson harmonics ($E_{J8}$ model). The parameters in each model are independently fitted from experimentally measured transmon frequencies~\cite{wang2025}. 
\rr{Because $n_g$ is not measured in this experiment, the simulated transition probabilities are averaged over 21 values of $n_g$ in the range $[0, 0.5]$. In \cref{sec: Floquet_Q4Q0}, we numerically investigate the dependence of ionization on $n_g$ and find that, as expected, the resonance condition of ionization depends only weakly on the offset charge for low-lying states of high-$E_J/E_C$ transmons.}
% \rr{The simulated transition probabilities are averaged over 21 values of $n_g$ in the range $[0, 0.5]$, and we expect our experiments to average over several offset-charge configurations ($10^4$ repetitions, 20 minutes total run time for each initial state). \bd{Empirical evidence for sufficient averaging is discussed in \cref{sec: t1_correlation}, where we show that the ionization probability remains fairly stable when the ionization experiment is repeated. Moreover, in} \cref{sec: Floquet_Q4Q0}, we investigate the dependence of ionization on $n_g$ and find that, as expected, the resonance condition of ionization only weakly depends on the offset charge for high-$E_J/E_C$ transmons.} 
The simulated transmon populations for the $E_{J1}$ and $E_{J8}$ models are respectively shown as solid and dashed lines in \cref{fig: 3_comparisons}(a,b,d,e). 

%% This paragraph tells what theoretical data/results are shown.
Both the predicted critical photon number and the amount of population transfer beyond the ionization point are well captured by the simulation using the $E_{J8}$ model, which validates the effectiveness of the aforementioned semiclassical model. We find that for all prepared initial states, the $E_{J8}$ model shows better agreement with experimental data than the $E_{J1}$ model. This is because the resonance conditions for ionization are sensitive to the transmon transition frequencies: Including additional Josephson harmonics in the transmon Hamiltonian plays a key role in accurately predicting the frequency of highly excited states and, thus, the occurrence of transmon ionization.

%% This paragraph tells what the Floquet branch is and how to calculate it.
The ionization process can also be understood via the more intuitive and more computationally efficient Floquet branch analysis~\cite{dumas2024}. Because the average photon number $\bar n_r(t)$ and phase $\phi(t)$ in \cref{eq:driven_harmonic_transmon_hamiltonian} vary slowly on the timescale of the drive period $T_d=2\pi/\omega_d$, the transmon Hamiltonian is approximately periodic on short timescales. As a result, ionization is determined by resonances in the Floquet spectrum associated with the instantaneously periodic Hamiltonian. To obtain this Floquet spectrum, we choose linearly spaced effective transmon drive amplitudes $2g\sqrt{\bar n_r}$ in steps of $2\pi \times \qty{100}{\kHz}$. For each constant drive amplitude, we calculate the Floquet modes and quasienergies by solving the eigenvalue problem of the propagator $U(T_d, 0)$ for \cref{eq:driven_harmonic_transmon_hamiltonian}. At $\bar n_{r} = 0$, the result coincides with the bare transmon eigenstates and eigenenergies, from where we sort other Floquet modes and quasienergies at higher amplitudes into a ``Floquet branch'' for each bare transmon state \cite{dumas2024}.

%% This paragraph explains the result of Floquet analysis.
We show the normalized quasienergies $\epsilon_j/\omega_d$ of the Floquet branches for $\omega_d=\omega_{r, \ket{0}}$ in \cref{fig: 3_comparisons}(c) and $\omega_d=\omega_{r, \ket{1}}$ in \cref{fig: 3_comparisons}(f), respectively, highlighting in color the most relevant Floquet branches for the $\ket{0} \leftrightarrow \ket{6}$ and $\ket{1} \leftrightarrow \ket{7}$ resonances. The Floquet branch for a given initial state shows avoided crossings with other branches as $\bar n_r$ increases. These avoided crossings indicate the multiphoton transitions responsible for ionization (the quasienergies are defined only modulo $\omega_d$). In general, there are multiple avoided crossings associated with various pairs of Floquet branches. However, we find that the positions of the largest avoided crossings, as well as the branches they involve, are consistent with those observed in the experiments and the dynamical simulations. Once ionization has occurred at these dominant avoided crossings, any number of other avoided crossings involving any of the populated branches can become relevant. This can occur during both the ramp-up and the ring-down phases of the readout pulse sequence.

%%%%%%%%%%%%%%%%%%%%%%%%%%%%%%%%%%%%%%%%
\section{\label{sec: LandauZener}Landau-Zener transitions}

\begin{figure*}[!t]
\includegraphics{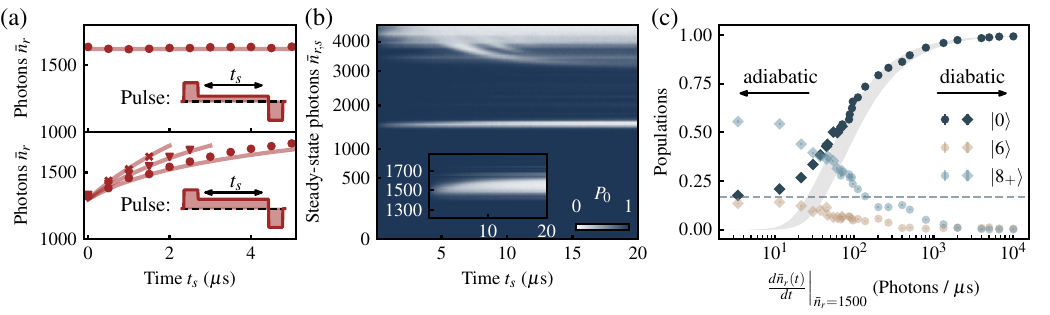}
\caption{
\label{fig: 4_LandauZener} 
Landau-Zener transitions. (a) The measured photon numbers (red dots) of the steady-state sequence (top) and Landau-Zener sequence (bottom) for different steady-state times $t_s$. The data agree well with the numerical prediction (red lines), which uses parameters extracted from independent measurements. The transmon is prepared in $\ket{0}$ at the beginning of the sequence. The insets show the envelopes of the shaped stimulation pulses, each of which includes three segments: ramp-up, steady state, and ramp-down. In the Landau-Zener sequence, the amplitude during $t_s$ is intentionally increased above the amplitude used for the steady-state sequence to drive the resonator from $\bar n_{r, i}$ to $\bar n_{r, f}$. The pulse is followed by a \qty{6}{\us} ring-down and an end-sequence measurement. We show three sequences with different Landau-Zener speeds identified by different symbols in the bottom panel. (b) The measured population of $\ket{0}$ from the end-sequence measurement for different steady-state durations $t_s$ and photon numbers $\bar n_{r, s}$. We observe a critical photon number \rr{$n_{r, \rm crit}$} for transmon ionization at around 1500 photons (see inset). Above 3000 photons, more resonances appear, while our pulse-shaping method fails to stabilize the photon number due to the higher-order nonlinearities of the resonator. (c) The measured populations of Landau-Zener experiments with different Landau-Zener speeds. The slope $d\bar n_r(t)/dt$ is controlled by changing the duration $t_s$ with fixed $\bar n_{r, i}=1300$, $\bar n_{r, f}=1700$ (circles), or by changing the difference $\bar n_{r, f}-\bar n_{r, i}$ with fixed $t_s=\qty{10}{\us}$ \rr{(diamonds)}. The horizontal gray dashed line shows the remaining ground-state population $P_0$ measured in the steady-state experiment \rr{at $t_s = \qty{10}{\us}$}, which has $\bar n_{r, f}-\bar n_{r, i} = 0$. The gray shaded area shows the range of theoretical predictions for 51 evenly spaced values of offset charge $n_g$. An adiabatic process results in more ionized population.
}
\end{figure*}

The avoided crossings in the Floquet quasienergies suggest that transmon ionization is a Landau-Zener-type transition~\cite{shillito2022, dumas2024,breuer1989}. A Landau-Zener process describes the dynamics of a two-level system evolving under a time-dependent Hamiltonian, where an external control field sweeps the system through an avoided crossing in its spectrum~\cite{Grifoni1998, ivakhnenko2023}. For transmon ionization, the photon number $\bar n_r$ plays the role of this control field, and the adiabaticity of the transition is determined by the speed at which $\bar n_r$ crosses the critical photon number \rr{$n_{r, \rm crit}$}~\cite{shillito2022, dumas2024}. If the transmon is prepared in an eigenstate and traverses an avoided crossing diabatically, ionization does not occur. Conversely, an adiabatic passage results in ionization. 

In the experiments shown in \cref{fig: 2_experiments}, the resonator never reached a steady state, and the adiabaticity was not carefully controlled by the square stimulation pulse. To remedy this, we investigate the Landau-Zener physics of transmon ionization using pulse-shaping techniques applied to transmon $Q_B$, which has 9 resolvable states\rr{; see \cref{sec: device_params} for its parameters}. We control the dynamics of the photon number using two different sequences, which we refer to as the steady-state sequence and the Landau-Zener sequence. The measured photon numbers (dots) and the numerical predictions (line) for the two sequences are shown in \cref{fig: 4_LandauZener}(a), where all parameters used in the numerical simulations are extracted from independent measurements. Details of the pulse calibration can be found in \cref{sec: pulses}.

In the steady-state sequence, the transmon is initially prepared in $\ket{0}$, followed by a stimulation pulse with three segments; see the top panel of \cref{fig: 4_LandauZener}(a). The first segment is a \qty{40}{\ns} ramp-up to rapidly bring the resonator from the vacuum state to the desired photon number $\bar n_{r,s}$. The second segment holds the resonator in its steady state with variable duration $t_s$. The third segment is a \qty{40}{\ns} ramp-down to actively empty the resonator. The frequency of the stimulation pulse is detuned from $\omega_{r, \ket{0}}$ by the expected Kerr shift $\bar n_{r,s} K_{r, \ket{0}}$ to compensate for the Kerr effect in the steady state. The short ramping time ensures strong diabaticity during ramping, making the transmon ionization most likely to happen during the steady-state segment. We end the sequence with a \qty{6}{\us} free ring-down time to fully empty the resonator, followed by a measurement.

The results of the steady-state experiment for different durations $t_s$ and photon numbers $\bar n_{r,s}$ are shown in \cref{fig: 4_LandauZener}(b). At low photon numbers, the system remains below \rr{$n_{r, \rm crit}$}, and the transmon stays in its initial state $\ket{0}$. At \rr{$n_{r, \rm crit} \approx 1500$}, however, the transmon ionizes, with a longer $t_s$ resulting in a lower ground-state population. Following the same method as in \cref{sec: experiments}, we identify the state $\ket{6}$ as one of the post-ionization states, and the transmon is further ionized to higher excited states; see \cref{sec: full_fig4b}. Above \rr{$n_{r, \rm crit}$}, there exists a range of photon numbers where the transition is suppressed again. This is because the system diabatically crosses the resonance at \rr{$n_{r, \rm crit}$} during the ramping segments while remaining far from other avoided crossings responsible for ionization during the steady-state segment. A similar phenomenon was observed in Ref.~\cite{sank2016}. At $\bar n_{r,s} > 3000$, there is again a reduction in the ground-state population. This occurs because the higher-order nonlinearities of the resonator become too strong for our pulse to stabilize the photon number for a long time while more avoided crossings appear. As a result, the resonator photon number sweeps through strong resonances, and most of the population eventually transfers from $\ket{0}$ to higher excited states.

Having identified the critical photon number \rr{$n_{r, \rm crit}$} for ionization, we next perform Landau-Zener experiments. The Landau-Zener sequence differs from the steady-state sequence in two ways; see the bottom panel of \cref{fig: 4_LandauZener}(a). First, after rapidly filling $\bar n_{r, i}$ photons into the resonator, the pulse amplitude of the steady-state segment is adjusted so that the photon number $\bar n_r(t)$ increases from $\bar n_{r, i}$ to $\bar n_{r, f}$ during the time $t_s$. As a result, the slope $d \bar n_r(t)/dt$ near the avoided crossing can be controlled by changing either $t_s$ or $\bar n_{r, f}-\bar n_{r, i}$. Second, for each slope, the pulse frequency is numerically optimized to compensate for the average Kerr effect during all segments; see \cref{sec: LZ_pulse}. The Landau-Zener speed near the avoided crossing is thus determined by the slope $d \bar n_r(t)/dt$ close to the critical photon number \rr{$n_{r, \rm crit}$}, with a flat (steep) slope corresponding to an adiabatic (diabatic) process. To reach a wide range of Landau-Zener speeds, we use two different parameterizations of the slope. In the diabatic regime, we fix $\bar n_{r,i}=1300$ and $\bar n_{r,f}=1700$, while sweeping the duration $t_s$ from \qty{40}{ns} to \qty{13}{\us}. In the adiabatic regime, we fix $t_s=\qty{10}{\us}$ and sweep the difference $\bar n_{r, f}-\bar n_{r, i}$ while keeping the mean photon number constant, $(\bar n_{r, f}+\bar n_{r, i}) / 2 = 1500$. \rr{The free ring-down time at the end of the Landau-Zener sequence is also reduced to \qty{1}{\us} to mitigate the effect of relaxation.} The measured final populations of state $\ket{0}$, state $\ket{6}$, and the combined populations of states $\ket{8}$ or higher are shown in \cref{fig: 4_LandauZener}(c) for both the diabatic regime (circles) and the adiabatic regime (diamonds). More details about the Landau-Zener sequence and its parameterizations can be found in \cref{sec: LZ_pulse}. 

The general trend of our experimental results matches the expectation of Landau-Zener physics, where a more adiabatic transition causes more population to be ionized. The remaining population $P_0$ in the adiabatic regime approaches \rr{a lower limit corresponding to} the result measured in the steady-state experiment, shown as the gray dashed line in \cref{fig: 4_LandauZener}(c). To obtain a quantitative comparison between the experiment and the theory, we use Floquet branch analysis. Similar to the method used in \cref{sec: comparisons}, we first calculate the Floquet spectrum at different photon numbers and sort them into Floquet branches. The related avoided crossings are identified by diabatically following the ground-state Floquet branch up to $\bar n_r = 2100$. We then select the avoided crossing with the largest gap $\Delta_{\rm{ac}}$ and compute the Landau-Zener diabatic transition probability 
\begin{equation}
P_{\rm{LZ}}=\exp(-\pi \Delta_{\rm{ac}}^2/2v).
\label{eq: FLZ}
\end{equation}
Here, the speed $v$ describes the rate of change of the gap near the avoided crossing and can be obtained from the Floquet branches and the experimentally measured $d \bar n_r(t)/dt$; see \cref{sec: LZ_speed} for more details. This transition probability corresponds to the ground-state population $P_0$ after the sequence. Because the Floquet spectrum depends on the offset charge $n_g$, the gray shaded area in \cref{fig: 4_LandauZener}(c) shows a range of probabilities calculated by choosing 51 different values of $n_g$. We find that the theoretical prediction reproduces the experimental trend very well, especially in the diabatic regime. In the adiabatic regime, the theoretical prediction increasingly deviates from the observed values as the Landau-Zener speed is lowered, because the time spent near the avoided crossing becomes increasingly comparable to the relaxation time of the excited states involved in ionization. The fully coherent Landau-Zener formula is thus not expected to accurately describe the transferred population.

\rr{
%%%%%%%%%%%%%%%%%%%%%%%%%%%%%%%%%%%%%%%%
\section{\label{sec: typical_transmons}Validation of the offset charge dependence for typical transmons}
}

\begin{figure*}[!t]
\includegraphics{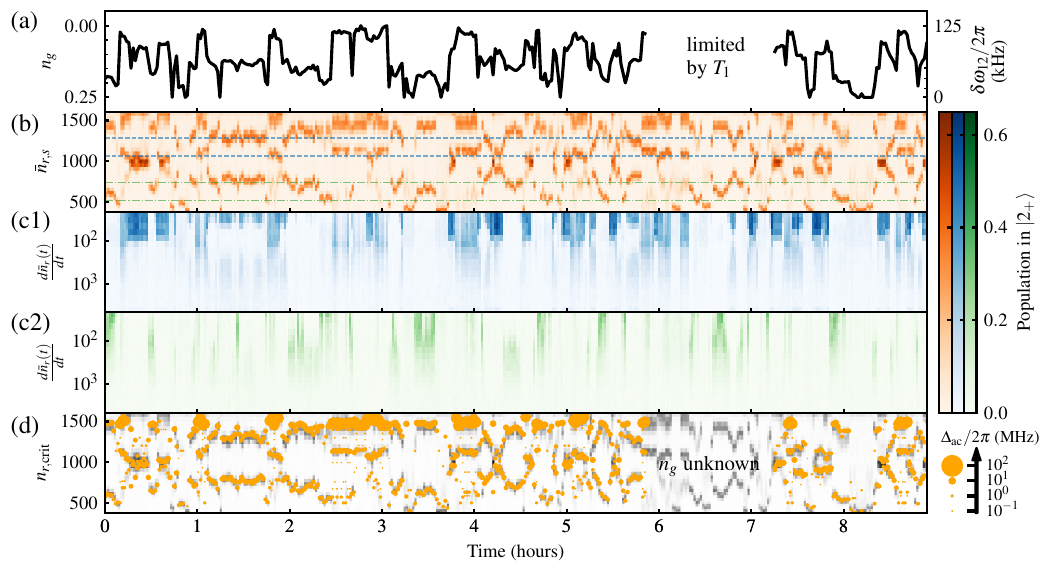}
\caption{
\label{fig: 5_timetrace} 
Time traces of the results of the repeated interleaved experiments on $Q_C$. (a) $n_g$ measurements through Ramsey experiments in the $\{\ket{1},\ket{2}\}$ subspace. Each value of $n_g$ is extracted by fitting two frequency components in a Ramsey experiment, corresponding to the odd and even charge parity of the transmon. Each Ramsey experiment takes 31 seconds. (b) $n_{r,\rm crit}$ measurements through steady-state experiments with fixed duration $t_s = \qty{16}{\us}$ and varying photon numbers $\bar n_{r,s}$. The transmon is prepared in $\ket{0}$ at the beginning of the sequence, and we show the ionized population $P_{2_+}$ here. The colored dashed-dotted lines encompass the \bd{photon number ranges of} two Landau-Zener experiments. Each steady-state experiment takes 22 seconds. (c) Landau-Zener experiments with different Landau-Zener speeds. We fixed the initial and final photon numbers as $\bar n_{r, i} = 1065$, $\bar n_{r, f} = 1286$ in (c1) and $\bar n_{r, i} = 519$, $\bar n_{r, f} = 735$ in (c2). The Landau-Zener speeds are controlled by changing $t_s$. A smaller value of $d\bar n_r(t)/dt$ corresponds to a more adiabatic process. Each Landau-Zener experiment takes 15 seconds. A single cycle of the interleaved experiments thus takes 83 seconds. (d) Theoretically predicted $n_{r, \rm crit}$. The values of $n_{r, \rm crit}$ are extracted by identifying the avoided crossings in the Floquet spectrum, and a larger marker represents a larger gap of the avoided crossing $\Delta_{\rm ac}$. For each time, we use experimentally measured $n_g$ to calculate the Floquet spectrum. The theoretical result is overlaid on the 2D data from panel (b) for comparison, with the latter shown in gray.
}
\end{figure*}

Having validated our experimental and numerical methods for highly excited states of high-$E_J/E_C$ transmons, next we focus on a typical transmon, $Q_C$, with $E_{J1}/E_C \approx 55$; see \cref{sec: device_params} for its full parameters. Ionization of a typical transmon differs from high-$E_J/E_C$ transmons in two ways~\cite{fechant2025}. First, the resulting states are more challenging to measure, as they are usually out of the potential well. Second, the critical photon number $n_{r, \rm crit}$ displays temporal fluctuations because the highly excited state of a typical transmon strongly depends on the offset charge $n_g$. Therefore, a quantitative comparison between theory and experiment needs to account for temporal fluctuations owing to charge noise.

\begin{figure}[!t]
\includegraphics{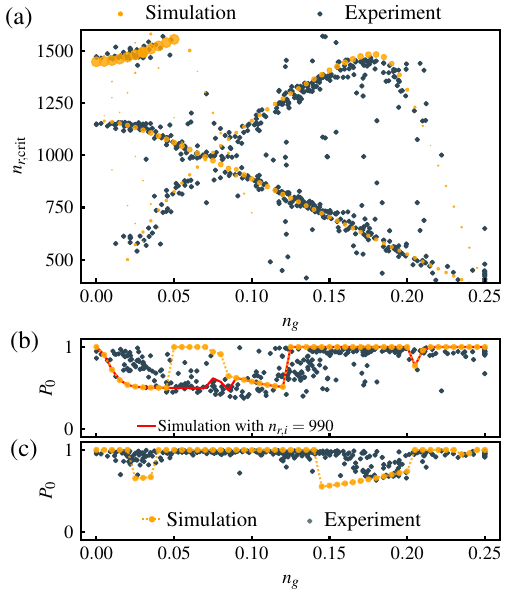}
\caption{
\label{fig: 6_ng} 
Comparing experimental and theoretical $n_g$ dependence of ionization on $Q_c$. (a) Critical photon numbers \bd{as a function of} $n_g$. \bd{The blue dots are the experimental values and the orange dots are the simulation}. The marker sizes of the simulation results are the same as in \cref{fig: 5_timetrace}(d). (b, c) Populations remaining in $\ket{0}$ in the Landau-Zener experiments with $\bar n_{r, i} = 1065$, $\bar n_{r, f} = 1286$ for (b) and $\bar n_{r, i} = 519$, $\bar n_{r, f} = 735$ for (c). The experimental populations (blue dots) correspond to the most adiabatic result, i.e., the smallest slope $d\bar n_r(t)/dt$. The simulated populations (orange dots) are calculated using the Landau-Zener formula, assuming the system crosses all avoided crossings in the range $[\bar n_{r, i}, \bar n_{r, f}]$. The red line in (b) shows \bd{the result of the simulation if we set $\bar n_{r, i} = 990$ instead}, which ensures that there is at least one avoided crossing in the selected range of \bd{photon numbers}.
}
\end{figure}

We characterize the time-dependent ionization of a typical transmon with four interleaved experiments over 8 hours to extract the offset charge $n_g$, the critical photon numbers $n_{r, \rm crit}$, and Landau-Zener dynamics across two different ranges of photon numbers. The experimental results are shown in \cref{fig: 5_timetrace}(a-d), and the comparison to theory is discussed in \cref{fig: 5_timetrace}(e) and \cref{fig: 6_ng}.

The first experiment, shown in \cref{fig: 5_timetrace}(a), is a Ramsey experiment in the $\{\ket{1},\ket{2}\}$ subspace to measure the slow drifts of $n_g$~\cite{serniak2018, wang2025}. The duration of each Ramsey experiment is 31 seconds, which is long compared to typical quasi-particle tunneling times and relatively short compared to the timescale of charge drifts~\cite{riste2013, serniak2018, serniak2019a, pan2022, connolly2024}. As a result, each Ramsey experiment shows a beating between two transition frequencies for a single $n_g$ configuration, corresponding to the even and odd charge parity of the transmon, and the value of $n_g$ can be extracted by fitting the difference of the two frequencies $\delta \omega_{12}$ for each experiment.
%; see \cref{sec: ng_conversion} for the conversion between $n_g$ and $\delta \omega_{12}$. 
We note that this method can only \bd{yield} a unique value of $n_g$ in \bd{the range} $[0,0.25]$, \bd{since $n_g$ and $0.5-n_g$ yield the same Ramsey frequencies}.

The second experiment, shown in \cref{fig: 5_timetrace}(b), measures $n_{r, \rm crit}$ by sweeping the steady-state photon number $\bar n_{r,s}$ analogous to \cref{fig: 4_LandauZener}(b). Here, we fix $t_s=\qty{16}{\us}$, and each steady-state experiment takes 22 seconds. The transmon is prepared in $\ket{0}$ at the beginning of the sequence, and we plot the ionized population $P_{2_+}$. Nonzero value of $P_{2_+}$ indicates the occurrence of ionization. In the photon range chosen here, multiple critical photon numbers $n_{r, \rm crit}$ are found because the system passes \bd{through} several multiphoton resonances between different pairs of states. Since the drift in $n_g$ causes changes in the transmon energy spectrum, the measured $n_{r, \rm crit}$ also fluctuates over time and correlates with $n_g$. This correlation can be seen by comparing the temporal positions of the features in panels (a) and (b).

The third experiment, shown in \cref{fig: 5_timetrace}(c1), is a Landau-Zener experiment analogous to \cref{fig: 4_LandauZener}(c). The transmon is also initially prepared in $\ket{0}$, and we plot the ionized population $P_{2_+}$ for different Landau-Zener speeds. Here, we use $\bar n_{r, i} = 1065$, $\bar n_{r, f} = 1286$, and the slope $d \bar n_r(t)/dt$ is controlled by only sweeping $t_s$. These values of photon numbers are chosen to represent a high photon number range and are corrected from the measured values considering the higher-order nonlinearities of the resonator; see \cref{sec:Kerr_effect}. We \bd{find} that, when at least \bd{one of the resonances observed in \cref{fig: 5_timetrace}(b)} falls into this photon range \bd{(dashed blue lines)}, we observe clear Landau-Zener dynamics: more populations are ionized in a more adiabatic process. When no resonance is in this range, no ionization occurs despite the adiabaticity.

The fourth experiment, shown in \cref{fig: 5_timetrace}(c2), is similar to the third one but with $\bar n_{r, i} = 519$, $\bar n_{r, f} = 735$ to \bd{probe} a \bd{lower} photon number range. \bd{We again find that the occurrence of Landau-Zener dynamics is correlated with the observation of resonances in this photon number range in \cref{fig: 5_timetrace}(b) (dashed-dotted green lines).} Each Landau-Zener experiment takes 15 seconds, and a single cycle of the interleaved experiments thus takes 83 seconds. The fast experimental speed ensures sufficient time resolution between different configurations of $n_g$ and $n_{r, \rm crit}$.

The numerical simulations shown in \cref{fig: 5_timetrace}(d) further confirm the correlation between $n_g$ and $n_{r, \rm crit}$. Here, we simulate the time dependence of the $n_{r, \rm crit}$. The values of $n_{r, \rm crit}$ are extracted from two Floquet spectra, which are calculated using instantaneous values of $n_g$ and $0.5-n_g$ measured in the Ramsey experiments. The simulated results \bd{are in excellent agreement} with the experimental time trace in \cref{fig: 5_timetrace}(b).

For a more quantitative comparison between the theoretical predictions and the experimentally observed $n_g$ dependence of $n_{r, \rm crit}$, we plot the same data as a function of $n_g$ instead of time in \cref{fig: 6_ng}. In panel (a), we \bd{overlay} the simulated \bd{values of $n_{r, \rm crit}$} (orange dots) and \bd{the} experimentally measured \bd{values of} $n_{r, \rm crit}$ (blue dots). The agreement is excellent except for a few outliers. The presence of outliers in the experimental results stems from the finite duration of each experiment and the delay between the $n_g$ measurement and the $n_{r, \rm crit}$ measurement. In other words, the actual value of $n_g$ during each $n_{r, \rm crit}$ measurement may fluctuate away from the value measured in \bd{the preceding} Ramsey experiment. Increasing the measurement speed or actively controlling $n_g$ \bd{would} help reduce this effect~\cite{fechant2025}. \rrr{Interestingly, we note that the good agreement between measurement and theory in \cref{fig: 6_ng}(a) provides a way to infer $n_g$ from the critical photon number, which could be useful when conventional offset-charge calibration methods fail.}

In panels (b) and (c), we show the results of the Landau-Zener experiments for different values of $n_g$ (blue dots). The experimental values are reproduced from the first row in \cref{fig: 5_timetrace}(c1) and (c2), corresponding to the most adiabatic process in each Landau-Zener experiment. The simulation results (orange dots) are calculated using \cref{eq: FLZ} assuming the same values of $\bar n_{r, i}$ and $\bar n_{r, f}$ as those measured in the experiments. When there are multiple avoided crossings in this photon range, we assume \bd{that} the system crosses all of them and extract the final diabatic transition probability as the ground state population $P_0$. The simulations agree with experiments on the general trend except for $n_g$ in the range $[0.05, 0.08]$ in panel (b). We attribute the discrepancy to small errors in calibrating the stimulation pulses of the Landau-Zener sequence. Moreover, the Floquet analysis we used here ignores the quantum fluctuation of the coherent state. As a result of these two factors, the photon numbers achieved in the experiment can span a broader range than those used in the simulation. If we set $\bar n_{r, i}=990$ in the simulation (red line) to ensure that there is at least one avoided crossing in the selected range of photon numbers, then \bd{we find} a better agreement.

\rrr{
We note that a recently published work, Ref.~\cite{fechant2025}, studies the offset-charge dependence of ionization by actively calibrating $n_g$ in flux-tunable low-$E_J/E_C$ transmons. Our work, by contrast, measures the offset charge $n_g$ in a time-resolved manner, which helps reveal the temporal fluctuations of the characteristic features of transmon ionization. Moreoever, using our pulse-shaping technique, we are able to selectively probe multiple critical photon numbers $n_{r, \rm crit}$ and their associated Landau-Zener transition probabilities as a function of $n_g$. Both works show the $n_g$ dependence of the critical photon number, provide verification of the semiclassical model introduced in Ref.~\cite{dumas2024}, and emphasize the importance of including transmon harmonics for accurately modeling the ionization process.
}

%%%%%%%%%%%%%%%%%%%%%%%%%%%%%%%%%%%%%%%%
\section{\label{sec: discussion}Discussion and outlook}

The transmon ionization is a key bottleneck for achieving fast, high-fidelity, high-QNDness measurement, which is necessary for many tasks in quantum information processing. In this work, we studied the excited-state dynamics of ionization in high-$E_J/E_C$ transmons. The deep potentials of such transmons enable control and readout of a large number of excited states, which allows us to observe the rich dynamics of ionization. As an example, for our parameters, we identify $\ket{7}$ as one of the final states when the transmon is prepared in $\ket{1}$. This identification is further verified by investigating the reverse ``deionization" process from $\ket{7}$ to $\ket{1}$. The photon numbers at which the transitions happen are consistent with each other, indicating the resonant nature of such processes.

Our work further validates the effectiveness of the driven transmon model and Floquet analysis. The comparison between the experimental results and the dynamical simulations shows excellent agreement for both the critical photon numbers and the ionized population. The Schrödinger equation simulation captures the majority of the experimental features. Additional effects, such as measurement-induced decay and dephasing, require additional consideration, which we leave for future work. Our results also highlight the importance of Josephson harmonics for an accurate prediction of the transmon spectrum. Combined with computationally efficient Floquet analysis, ionization could be mitigated by optimizing the transmon and resonator parameters, such as the transition frequencies and the coupling strength, so that the threshold of ionization is increased. This threshold informs the maximum allowable photon number during readout, as shown in Ref.~\cite{bengtsson2024}, which helps avoid unwanted transitions.

The Landau-Zener physics is another strong evidence
of the two-level resonance. Using a pulse-shaping method, we demonstrate precise control of the photon number in the resonator, which allows us to pass through an avoided crossing over a wide range of adiabaticity. Our experimental results agree with the theoretical prediction that a more adiabatic process yields more population transfer. \rr{The pulse shaping method also allows us to verify the offset charge dependence of ionization for a typical transmon.} An intriguing question to answer in the future is whether a high-power QND readout is achievable by crossing the resonance diabatically. Moreover, the reported population in \cref{fig: 4_LandauZener}(c) corresponds to the transition probability for a single passage through the avoided crossing, and it could be possible to observe Landau-Zener-Stückelberg interference upon a double passage in future work.

Although ionization is often discussed in the context of qubit readout, related challenges can become more significant when using a transmon as a high-dimensional qudit~\cite{wang2025, champion2025}, since higher excited states introduce additional resonance conditions in the spectrum. An example is shown in \cref{fig: 2_experiments}(d), where population transfer between $\ket{7}$ and $\ket{9_+}$ occurs earlier than between $\ket{7}$ and $\ket{1}$. In addition, multitone readout—typically required for qudits~\cite{chen2023, wang2025}—leads to more complex ionization dynamics that cannot be captured by modeling the transmon as being driven by a single periodic tone. These observations highlight the need for careful consideration of readout pulse parameters in qudit applications.

The ability to control higher-energy levels of the transmon may also help in the investigation of ionization in alternative readout approaches, such as longitudinal readout~\cite{didier2015,chapple2025a} and balanced cross-Kerr readout~\cite{chapple2025b}. Moreover, although this work focuses on measurement-induced effects, similar effects are expected to arise in other contexts where the essential ingredients for ionization are present---namely, strong drives and nonlinearity--- such as parametric gates, qubit reset, and quantum state stabilization. Since these scenarios often involve transmon-like circuits, we expect that our work will provide new insights into the effect of strong drives on superconducting quantum circuits beyond readout.

%%%%%%%%%%%%%%%%%%%%%%%%%%%%%%%%%%%%%%%%
\section*{\label{sec: acknowledgement}Acknowledgment}
We thank Alexander McDonald, Crist\'{o}bal Lled\'{o}, and Marie Fr\'{e}d\'{e}rique Dumas for fruitful discussions. We thank Rayleigh William Parker for assistance in designing the experimental sample.

This work is supported by a collaboration between the US DOE and other Agencies. This material is based upon work supported by the U.S. Department of Energy, Office of Science, National Quantum Information Science Research Centers, Quantum Systems Accelerator. Additional support is acknowledged from Air Force Office of Scientific Research Grant No. FA9550-23-1-0121, NSERC, the Ministère de l’Économie et de l’Innovation du Québec, and the Canada First Research Excellence Fund. Devices used in this work were fabricated and provided by the Superconducting Qubits at Lincoln Laboratory (SQUILL) Foundry at MIT Lincoln Laboratory, with funding from the Laboratory for Physical Sciences (LPS) Qubit Collaboratory. The traveling-wave parametric amplifier (TWPA) used in this experiment was provided by IARPA and Lincoln Labs.

%%%%%%%%%%%%%%%%%%%%%%%%%%%%%%%%%%%%%%%%
\section*{\label{sec: data_availability}Data Availability}
The data that support the findings of this article are openly available~\cite{wang_data_2026}.

%%%%%%%%%%%%%%%%%%%%%%%%%%%%%%%%%%%%%%%%
\appendix

%%%%%%%%%%%%%%%%%%%%%%%%%%%%%%%%%%%%%%%%
\section{\label{sec: device_params}Device parameters and characterization}

\begin{figure}[!h]
\includegraphics{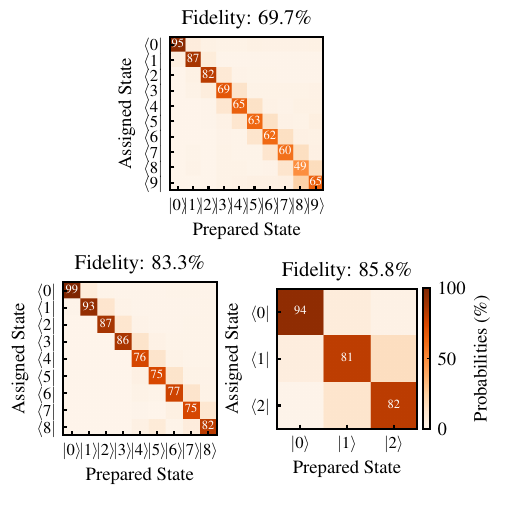}
\caption{
\label{fig: S8_ROfidelities} 
\rr{Readout assignment matrix for $Q_A$ (top), $Q_B$ (bottom left), and $Q_C$ (bottom right).}
}
\end{figure}

The device parameters used in this work are shown in \cref{tab: device_params}. \rr{$Q_A$ and $Q_B$ correspond to $Q_5$ and $Q_4$ of Ref.~\cite{wang2025}, and $Q_C$ is on the same chip but was not used in that reference. The length of the control pulse used to drive each adjacent transition is \qty{40}{\ns}. The errors in state preparation and single-qubit gates are on the order of $10^{-3}$, typically limited by the anharmonicity and the decoherence of the transmon~\cite{wang2025}. The readout assignment matrices and fidelities of the three transmons are shown in \cref{fig: S8_ROfidelities}. \Cref{tab: coherence} shows the coherence time, where the values of $Q_A$ and $Q_B$ are reproduced from Ref.~\cite{wang2025}.} Due to the thermal cycles of the sample, the parameters of the two transmons are slightly changed compared to those reported in Ref.~\cite{wang2025}, \rr{and the updated relaxation times of the $\ket{0} \leftrightarrow \ket{1}$ transition are $T_1=\qty{49(9)}{\us}$ for $Q_A$ and $T_1=\qty{28(9)}{\us}$ for $Q_B$. We note that the coherence time in Ramsey experiments of $Q_A$ became significantly shorter ($T_{2R} \sim \qty{2}{\us}$) after a thermal cycle, which motivated us to use $Q_B$ for further study of the Landau-Zener physics.} The control and stimulation pulses are generated using 16-bit DACs in the Qblox QCM-RF module, and the readout signals are generated and detected using the Qblox QRM-RF module.

\begin{table*}[!t]
\caption{\label{tab: device_params}Device parameters.}
\begin{ruledtabular}
\begin{tabular}{cccc}

Device & $Q_A$ & $Q_B$ & $Q_C$  \\ 
Usage  & \cref{sec: experiments,sec: comparisons} & \cref{sec: LandauZener} & \cref{sec: typical_transmons} \\

\hline

First anharmonicity $\alpha_1/2\pi = f_{12} - f_{01}$ (MHz) & -104 & -114 & -236 \\ 
0-1 transition frequency $\omega_{01}/2\pi$ (GHz) & 4.8817 & 4.8334 & 4.1968 \\
1-2 transition frequency $\omega_{12}/2\pi$ (GHz) & 4.7778 & 4.7198 & 3.9605 \\
2-3 transition frequency $\omega_{23}/2\pi$ (GHz) & 4.6694 & 4.6007 & - \\
3-4 transition frequency $\omega_{34}/2\pi$ (GHz) & 4.5557 & 4.4754 & - \\
4-5 transition frequency $\omega_{45}/2\pi$ (GHz) & 4.4361 & 4.3428 & - \\
5-6 transition frequency $\omega_{56}/2\pi$ (GHz) & 4.3098 & 4.2015 & - \\
6-7 transition frequency $\omega_{67}/2\pi$ (GHz) & 4.1753 & 4.0497 & - \\
7-8 transition frequency $\omega_{78}/2\pi$ (GHz) & 4.0310 & 3.8848 & - \\
8-9 transition frequency $\omega_{89}/2\pi$ (GHz) & 3.8746 & - & - \\

\hline

Resonator frequency when transmon is at $\ket{0}$ $\omega_{r, \ket{0}}/2\pi$ (GHz) & 6.470366 & 6.415708 & 6.311833 \\

Dispersive shift $(\omega_{r, \ket{1}} - \omega_{r, \ket{0}})/2\pi$ (kHz) & -249 & -205 & -267 \\

Resonator linewidth $\kappa/2\pi$ (kHz) & 105 & 127 & 129 \\

Josephson energy $E_{J1}/h$ (GHz) \footnote{Parameters here are estimated by $E_{J8}$ model for $Q_A$, $Q_B$ and $E_{J2}$ model for $Q_C$.} & 29.7 & 27.1 & 11.6 \\

Charging energy $E_C/h$ (GHz) \footnotemark[1] & 0.108 & 0.116 & 0.212 \\

$E_{J1}/E_C$ \footnotemark[1] & 275 & 235 & 55 \\ 
Transmon-resonator coupling strength $g/2\pi$ (MHz) \footnotemark[1] & 31.0 & 26.8 & 39.5\\

\hline
\end{tabular}
\end{ruledtabular}
\end{table*}

\begin{table*}[tp]
\rr{\caption{\label{tab: coherence} Coherence times $T_1/T_{2R}/T_{2E}$ measured from relaxation/Ramsey/Hahn Echo experiments, respectively, for all transmons in this work. The values for $Q_A$ and $Q_B$ are reproduced from Ref.~\cite{wang2025}, all in \unit{\us}.}}
\begin{ruledtabular}
\begin{tabular}{cccc}
Transition & $Q_A$ & $Q_B$ & $Q_C$ \\ 
\hline
$\ket{1} \rightarrow \ket{0}$ & 64(15)/85(31)/93(27) & 46(7)/52(11)/51(7) & 90(17)/63(16)/110(18) \\
$\ket{2} \rightarrow \ket{1}$ & 34(8)/51(19)/53(14)  & 25(4)/24(6)/37(5)  & 80(33)/41(17)/69(20)  \\
$\ket{3} \rightarrow \ket{2}$ & 24(5)/44(12)/45(10)  & 26(3)/15(4)/34(3)  & -/-/-  \\
$\ket{4} \rightarrow \ket{3}$ & 21(4)/39(11)/39(8)   & 14(3)/15(5)/25(7)  & -/-/-  \\
$\ket{5} \rightarrow \ket{4}$ & 17(3)/27(8)/32(7)    & 16(2)/48(12)/44(6) & -/-/-  \\
$\ket{6} \rightarrow \ket{5}$ & 14(3)/25(7)/26(6)    & 14(2)/28(10)/27(6) & -/-/-  \\
$\ket{7} \rightarrow \ket{6}$ & 13(3)/22(8)/24(6)    & 13(2)/20(9)/23(6)  & -/-/-  \\
$\ket{8} \rightarrow \ket{7}$ & 14(3)/21(7)/24(6)    & 11(2)/11(5)/20(4)  & -/-/-  \\
$\ket{9} \rightarrow \ket{8}$ & 13(2)/16(5)/22(5)    & -/-/-              & -/-/-  \\
\end{tabular}
\end{ruledtabular}
\end{table*}

%%%%%%%%%%%%%%%%%%%%%%%%%%%%%%%%%%%%%%%%
\rr{
\section{\label{sec: full_fig_2}Full population and calibration of $\bar n_{r, \rm max}$ for the square stimulation experiment}
}

\begin{figure}[!b]
\includegraphics{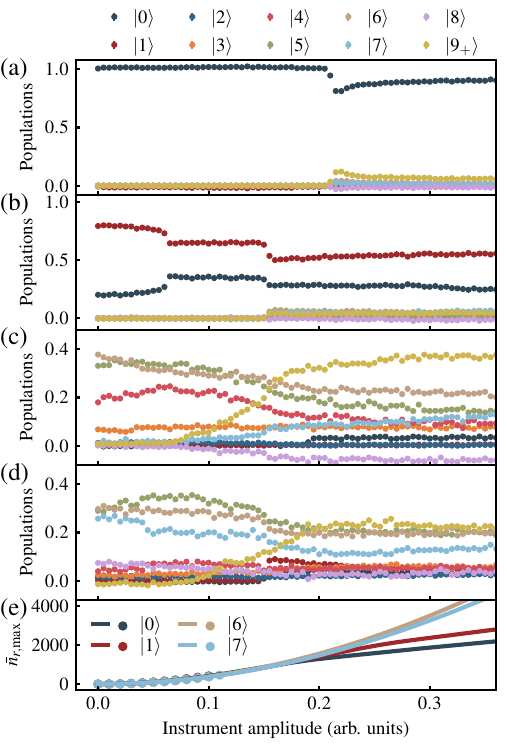}
\caption{
\label{fig: S1_photon_extrapolation}
\rr{(a-d) The populations of all states for the experiments reported in \cref{sec: experiments,sec: comparisons} as a function of the instrument amplitude. At the beginning of the sequence, the transmon is prepared in $\ket{0}$ for (a), $\ket{1}$ for (b), $\ket{6}$ for (c), and $\ket{7}$ for (d). (e) Measured $\bar n_{r, \rm{max}}$ (dots) and the extrapolation results (lines) as a function of amplitude}.
}
\end{figure}

In \cref{fig: 2_experiments}, we \bd{only show the population of some states} as a function of the maximum photon number for simplicity. Here we provide \bd{the populations of all states} as a function of the stimulation amplitudes in \cref{fig: S1_photon_extrapolation}(a-d) and give more details about the conversion between the instrument amplitude and the maximum photon number.

%%%%%%%%%%%%%%%%%%%%%%%%%%%%%%%%%%%%%%%%
\subsection{\label{sec:identify_target}Identification of the resulting states}

In \cref{sec: experiments}, we identify $\ket{1} \leftrightarrow \ket{7}$ as a pairwise ionization process. However, the population changes of $\ket{7}$ at $\bar n_{r, \rm max} \sim 880$ in \cref{fig: 2_experiments}(c) and (d) are both less significant compared to the population changes of $\ket{1}$. This is due to the relatively short relaxation time of $\ket{7}$ compared to $\ket{1}$\bd{. As a result,} the population is more likely to be transferred into lower states $\ket{6}$ and $\ket{5}$; see \cref{fig: S1_photon_extrapolation}(d) for \bd{the populations of all states}. \bd{Indeed, the expected} population change is recovered when the populations are summed into the qubit and leakage subspaces; see \cref{fig: 3_comparisons}. Meanwhile, at $\bar n_{r, \rm max} \sim 880$, the population drop of $\ket{7}$ in \cref{fig: 2_experiments}(d) is about $2.7\%$, which is less than the $4.4\%$ population increase of $\ket{7}$ in \cref{fig: 2_experiments}(c). \bd{This is because the stimulation frequencies for the two experiments are different. Indeed, this frequency difference is the cause of the discrepancy predicted by the theoretical results in} \cref{fig: 3_comparisons}(d, e). This effect, along with the decoherence and the readout errors, results in a \bd{lower and} unsharp population drop of $\ket{7}$ in \cref{fig: 2_experiments}(d). 

Here, one could argue that $\ket{6}$ is the destination, instead of $\ket{7}$, and \bd{that} the de-ionization process could occur after $\ket{7}$ decays to $\ket{6}$. This argument is disproved in \cref{fig: S1_photon_extrapolation}(c), where the transmon is prepared in $\ket{6}$ and \bd{yet} there is no population increase in $\ket{1}$ for any stimulation amplitude.

%%%%%%%%%%%%%%%%%%%%%%%%%%%%%%%%%%%%%%%%
\subsection{\label{sec:negative_populations}Mitigation of readout errors}

The populations reported in \cref{fig: 2_experiments,fig: S1_photon_extrapolation} are corrected populations. To obtain these populations, each readout signal is sampled, integrated, and assigned to a certain state. The populations of these states are then normalized and further corrected to mitigate readout errors due to limited readout assignment fidelity. We first give a brief review of our correction method and then explain \bd{possible reasons why it sometimes yields negative populations.}
%two possible reasons for negative populations.

For three-tone readout of $Q_A$ and $Q_B$, the integrated signal associated with each shot forms a six-dimensional $IQ$ vector $(I_a, Q_a, I_b, Q_b, I_c, Q_c)$. This vector is assigned to a transmon state using a Gaussian mixture model (GMM). The parameters of the GMM, such as its means and its covariance matrix, are fitted from an independent calibration. The calibration is done by preparing the transmon in each of its initial states and performing single-shot readout. The conditional probabilities of the calibration results form an assignment matrix $\mathcal{A}$, where $\mathcal{A}_{ij} = P(\text{assigned as } i|\text{prepared as }j)$. The GMM parameters and the assignment matrix are then used in the other experiments. In each experiment, the corrected populations are calculated by applying the inverse assignment matrix to the normalized populations, $P_{\rm{corr}}=\mathcal{A}^{-1} P_{\rm{norm}}$.

Although experiments are usually performed continuously and repetitively, the calibration for readout (and also \bd{for the} drive) is usually performed in a less frequent fashion. As a result, low-frequency noise in the system may cause fluctuations of the distribution of the $IQ$ vectors, i.e., the true parameters of the GMM. Such fluctuations are typically small and negligible in most experiments, but may become noticeable when the readout fidelity is limited, corresponding to strong overlap between the Gaussian mixtures. When applying the correction method mentioned above, the resulting populations may become negative. 
This fluctuation and the resulting negative population do not depend on the exact experimental sequence. For instance, in \cref{fig: 2_experiments}(d) and \cref{fig: S1_photon_extrapolation}(d), even at zero amplitude (corresponding to zero photons), where no physical stimulation pulse is applied, the population of $\ket{9_+}$ exhibits small negative values.

Apart from the \bd{fluctuations in the}
%stochastic distribution fluctuation of
GMM parameters, another reason for negative populations in our correction method is \bd{the population of uncalibrated states.}
%the emergence of additional Gaussian mixtures.
For instance, although the final states of $Q_A$ can be higher than $\ket{9}$, we only prepare and read out up to $\ket{9}$ during the readout calibration, and the higher states are collectively classified as $\ket{9_+}$, along with $\ket{9}$ itself. However, the $IQ$ distributions of these states are not exactly the same as $\ket{9}$. When a significant amount of transmon population is inside these states, the \bd{actual distribution of outcomes}
%actual measured results
may shift away from the calibrated distribution of $\ket{9}$. In our case, it causes even more $IQ$ vectors to be assigned as $\ket{9_+}$ than \bd{in} the calibration. As a result, after applying the correction method, the population in $\ket{9_+}$ can show a population greater than 1, and its closest state, $\ket{8}$, can show a negative population. As shown in \cref{fig: S1_photon_extrapolation}(c), the population increase in $\ket{9_+}$ and the negative population of $\ket{8}$ occur at the same instrument amplitude, which is likely due to the occupation of higher excited states for stronger stimulation amplitudes.

The two phenomena above would be strongly suppressed if the readout fidelity \bd{were} increased. A more frequent calibration with more readout shots would also help to address the first issue above. Instead of using the inverse assignment matrix, another way to apply readout correction is to treat the post-correction populations as unknown parameters and perform maximum likelihood estimation of these parameters while constraining them to the closed interval $[0, 1]$. We note that this alternative method may generate cutoff effects on the features of the experimental data.

%%%%%%%%%%%%%%%%%%%%%%%%%%%%%%%%%%%%%%%%
\subsection{\label{sec:extrapolation}Conversion between the instrument amplitude and the maximum photon number}

In \cref{sec: experiments} of the main text, we show the transmon populations as a function of the maximum photon number reached during the sequence. In practice, the experiments were performed by sweeping the instrument amplitude. For a small range of values, this amplitude is typically proportional to the drive amplitude $\varepsilon$ on the resonator. For a classical linear resonator initialized in the vacuum state and evolving under a resonant drive, the mean intra-resonator photon number $\bar n_r(t)$ at a given time $t$ is
\begin{equation}
\bar n_r(t) = \left( \frac{\varepsilon}{\kappa} \right)^2 (1 - e^{-\kappa t / 2})^2,
\label{eq: resonant_rampup_photons}
\end{equation}
where $\kappa$ is the decay rate of the resonator. \Cref{eq: resonant_rampup_photons} suggests the quadratic relationship $\bar n_r \propto \varepsilon^2$ for a fixed time $t$. However, the actual photon number may deviate from this quadratic behavior due to the Kerr effect. 

To find the accurate maximum mean photon number $\bar n_{r, \rm max}$, we first fit the conversion between the instrument amplitude and the measured $\bar n_{r, \rm max}$ at low power. We then extrapolate the photon numbers at higher power based on the driven Kerr resonator model explained in \cref{sec: pulses}. The fitting and extrapolations are repeated for different initial states $\ket{j}$ because the effective $\kappa$ weakly depends on the state of the transmon~\cite{blais2021}. The extrapolation results are shown in \cref{fig: S1_photon_extrapolation}(e), where the Kerr coefficient $K_{r, \ket{j}}$ is calculated from numerical diagonalization of the Hamiltonian.

%%%%%%%%%%%%%%%%%%%%%%%%%%%%%%%%%%%%%%%%
\section{\label{sec: QAQBswap}Interaction with neighbor transmon}

\begin{figure}[!h]
\includegraphics{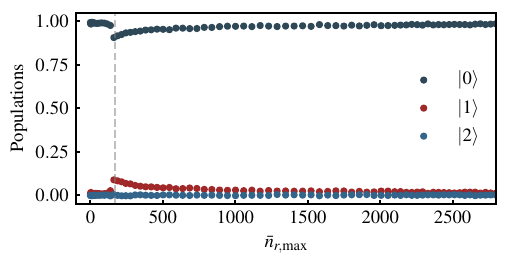}
\caption{
\label{fig:S2_QAQBswap} 
Eigenstate populations of $Q_B$ for the experimental pulse sequence used in Fig. 2(c) of the main text and in \cref{fig: S1_photon_extrapolation}(b).
}
\end{figure}

In \cref{fig: 2_experiments} of the main text and in \cref{fig: S1_photon_extrapolation}(b), we see a drop in the population of state $\ket{1}$ at $\bar n_{r,\rm{max}} \sim 170$. We find that it is due to a resonant population swap between $Q_A$ and $Q_B$, with $Q_B$ remaining idle during this experiment. The two transmons are fabricated on the same chip and designed to have negligible coupling with each other. However, when $Q_A$ is prepared in $\ket{1}$ and ac-Stark-shifted by \qty{-43}{\MHz}, it becomes resonant with the neighbor transmon $Q_B$, which has the frequency $\omega_{01} = 2\pi \times \qty{4.8380}{\GHz}$ at that thermal cycle. This resonance induces a $\ket{10} \leftrightarrow \ket{01}$ swap that causes a population drop in $Q_A$. In \cref{fig:S2_QAQBswap}, we show the measured populations of $Q_B$ under different stimulation powers on $Q_A$ using the same experimental sequence as in \cref{fig: 2_experiments}. The readout pulse on $Q_B$ is added immediately after the stimulation on $Q_A$ to probe transitions that occur during the ramp-up. This also has the benefit of reducing the effect of decay. We find a population peak at around 170 photons, which confirms the occurrence of the resonant swap. This mechanism is further confirmed by the absence of population transfer when preparing the ground state, in which case the swap between the qubits is not energetically possible. Finally, \cref{fig:S2_QAQBswap} also shows that the probability of a swap during stimulation becomes small at large photon numbers. This is because the associated resonance is crossed much more rapidly, reducing the probability of a Landau-Zener transition. Thus, at large drive powers, any significant population transfer to \rr{$Q_B$} must occur during the ring-down.

%%%%%%%%%%%%%%%%%%%%%%%%%%%%%%%%%%%%%%%%
\rr{
\section{\label{sec: t1_correlation}Population correlation with $T_1$}
}

In \cref{sec: comparisons} of the main text, we argue that the transmon relaxation mostly happens after the ionization and thus should have a relatively minor impact when comparing the experimental results with the theory, which excludes transmon relaxation. Here we provide supporting evidence for this argument.

\begin{figure}[t]
\includegraphics{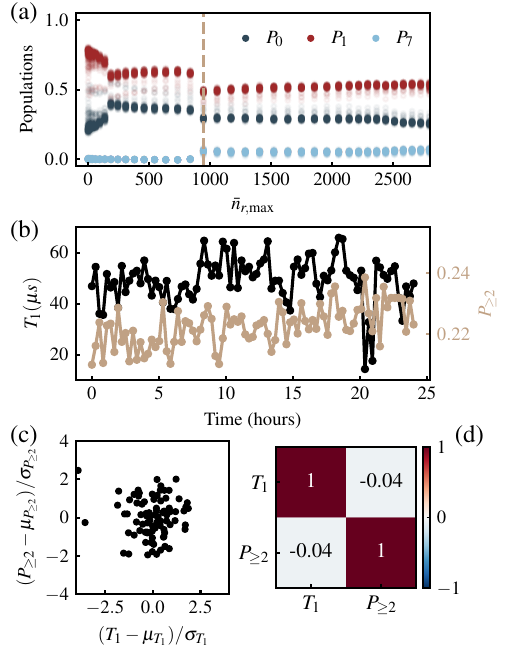}
\caption{
\label{fig: S7_T1correlation}
Interleaved ionization experiments and $T_1$ measurements. (a) Overlapped populations of states $\ket{0}$, $\ket{1}$, and $\ket{7}$ from 87 ionization experiments \bd{performed within} $\sim 24$ hours. The transmon is initially prepared in $\ket{1}$. (b) The time trace of ionized populations $P_{\geq 2}$ and measured $T_1$ in the $\ket{1} \leftrightarrow \ket{0}$ subspace. Here, the $P_{\geq 2}$ is evaluated at the photon number shown as the vertical dashed line in (a). (c) The scatter plot of the two variables in (b). Both variables are standardized. (d) The Pearson correlation coefficient matrix of the two variables.
}
\end{figure}

We repeat the ionization experiment in \cref{fig: 2_experiments}(c) for roughly 24 hours, where transmon $Q_A$ is prepared in $\ket{1}$ at the beginning of the sequence. We note that \bd{we use fewer values of the maximum photon number} $\bar n_{r, \rm{max}}$ \bd{than in \cref{fig: 2_experiments}}
%in the repeated experiments
\bd{to increase the} repetition rate. The populations of states $\ket{0}$, $\ket{1}$, and $\ket{7}$ are \bd{overlapped} in \cref{fig: S7_T1correlation}(a) \bd{for all 87 ionization experiments, showing that} the critical photon number is stable over time \bd{as is expected for the deep potential of transmon $Q_A$}.

These ionization experiments are interleaved with $T_1$ measurement in the $\ket{1} \leftrightarrow \ket{0}$ subspace. The resulting time trace of relaxation time $T_1$ and ionized population $P_{\geq 2}$ is shown in \cref{fig: S7_T1correlation}(b), where $P_{\geq 2}$ is evaluated at the critical photon number shown as the vertical dashed line in \cref{fig: S7_T1correlation}(a). We also show a scatter plot of two standardized variables and their Pearson correlation coefficient matrix in \cref{fig: S7_T1correlation}(c) and (d), respectively. These results exhibit a minor correlation between $T_1$ and $P_{\geq 2}$, which justifies \bd{excluding relaxation from our model.}
%our approach of using the model without relaxation. 

We note that, in general, the
% impact of
\bd{error made by} excluding relaxation in the theoretical model can strongly depend on the experimental sequence and on the parameters of the transmon-resonator system. For the results shown in \cref{sec: experiments,sec: comparisons}, the photon number peaks at the early stage of the sequence, see \cref{fig: 2_experiments}(b), which makes \bd{relaxation less relevant.}
%the result less affected by relaxation.
However, for the results shown in \cref{sec: LandauZener}, especially for the steady-state experiment and the Landau-Zener experiment in the adiabatic regime, the ionization can occur throughout the sequence, \bd{in which case relaxation could become more important.}
%which suggests the importance of including relaxation in the theoretical model.

%%%%%%%%%%%%%%%%%%%%%%%%%%%%%%%%%%%%%%%%
\section{\label{sec: pulses}Pulse-shaping and calibration for the steady-state and Landau-Zener experiments}

The steady-state experiment and the Landau-Zener experiment discussed in \cref{sec: LandauZener} require precise control over the state of the resonator. In this section, we explain our pulse-shaping method and give examples of numerical simulations and experimental calibration results.

%%%%%%%%%%%%%%%%%%%%%%%%%%%%%%%%%%%%%%%%
\subsection{\label{sec:EOM}Classical resonator model}

Consider a classical driven and damped Kerr resonator in a frame rotating at the drive frequency $\omega_d$. The equation of motion of its field $\alpha(t)$ is
\begin{equation}
\begin{split}
\dot{\alpha}(t) &= i \Delta \alpha(t) - iK_r|\alpha(t)|^2\alpha(t) -  \frac{\kappa}{2} \alpha(t) \\
&- i\frac{\varepsilon(t)}{2} e^{-i\phi_d},
\label{eq:EOM_Kerr_resonator}
\end{split}
\end{equation}
where $\Delta \equiv \omega_d - \omega_r$ is the drive-resonator detuning, $K_r$ is the Kerr coefficient of the resonator, and $\varepsilon(t)$ and $\phi_d$ are the amplitude and phase of the drive, respectively. When dispersively coupled to a transmon, the Kerr value is negative due to the negative anharmonicity of the transmon.

%%%%%%%%%%%%%%%%%%%%%%%%%%%%%%%%%%%%%%%%
\subsection{\label{sec:Linear_resonator}Linear resonator and three-segment pulse}

For a linear resonator ($K_r=0$) under a constant resonant drive [$\varepsilon(t)=\varepsilon$, $\Delta=0$], the solution of \cref{eq:EOM_Kerr_resonator} is
\begin{equation}
\alpha(t) = C e^{-\kappa t / 2} - i e^{-i\phi_d} \frac{\varepsilon}{\kappa},
\label{eq:solution_linear_rampup}
\end{equation}
where $C$ is an integral constant depending on the initial condition. If the resonator starts in the vacuum state, $\alpha(0)=0$, then
\begin{equation}
\alpha(t) = - i e^{-i\phi_d} \frac{\varepsilon}{\kappa} (1 - e^{-\kappa t / 2}),
\label{eq:solution_linear_rampup_vacuum}
\end{equation}
and the mean photon number reduces to \cref{eq: resonant_rampup_photons} with steady-state value $\bar n_r=|\alpha(t)|^2=\varepsilon^2/\kappa^2$. Because it will be useful below, we note that \cref{eq:solution_linear_rampup_vacuum} is expressed as a real function of time multiplied by a time-independent global phase.

Our goal is to construct a shaped pulse to drive the resonator such that: (a) the resonator is ramped up to its steady state as fast as possible, (b) the steady state is then stabilized for a long time, and (c) the resonator is rapidly ramped down to the vacuum state at the end. We denote the ramp-up, steady state, and ramp-down times as $t_{\uparrow}$, $t_s$, and $t_{\downarrow}$, during which the drive amplitudes $\varepsilon_{\uparrow}$, $\varepsilon_s$, and $\varepsilon_{\downarrow}$ are applied, respectively. The drive amplitudes remain constant inside each segment and thus form a step-wise pulse as shown in the inset of \cref{fig: 4_LandauZener}(a) in the main text. We emphasize that the sequence is followed by an additional free ring-down with time $t_{rd}$, which is different from the active ramp-down segment above. This ring-down is added to further ensure the resonator is empty before the final measurement.

To reach the steady state photon number $\varepsilon_s^2/\kappa^2$ in a given time $t_{\uparrow}$, the following equality must be satisfied
\begin{equation}
\bar n_r(t_\uparrow) 
= \left( \frac{\varepsilon_\uparrow}{\kappa} \right)^2 (1 - e^{-\kappa t_\uparrow / 2})^2 
= \frac{\varepsilon_s^2}{\kappa^2}.
\label{eq:rampup_photons}
\end{equation}
If the phase of the pulse is the same for all segments, then \cref{eq:rampup_photons} gives the relation
\begin{equation}
\frac{\varepsilon_\uparrow}{\varepsilon_s} = \frac{1}{1 - e^{-\kappa t_\uparrow / 2}} > 1.
\label{eq:rampup_amplitude}
\end{equation}
As an example, the resonator $R_4$ used in this work has $\kappa = 2\pi \times \qty{127}{\kHz}$, which requires an amplitude ratio $\varepsilon_\uparrow/\varepsilon_s \approx 63.3$ for a ramp-up time $t_{\uparrow}=\qty{40}{\ns}$. 

Similarly, the ramp-down amplitude $\varepsilon_\downarrow$ should satisfy
\begin{equation}
\begin{split}
\frac{\varepsilon_\downarrow}{\varepsilon_s} 
&= \frac{e^{-\kappa t_\downarrow / 2}}{e^{-\kappa t_\downarrow / 2} - 1} \\
&= 1 - \frac{\varepsilon_\uparrow}{\varepsilon_s} < 0, \quad \text{if } t_\uparrow=t_\downarrow.
\label{eq:rampdown_amplitude}
\end{split}
\end{equation}

%%%%%%%%%%%%%%%%%%%%%%%%%%%%%%%%%%%%%%%%
\subsection{\label{sec: Kerr_resonator}Kerr resonator and detuning}

The pulse introduced in \cref{sec:Linear_resonator} can stabilize linear resonators because the drive term in \cref{eq:EOM_Kerr_resonator} balances the damping term. However, the Kerr effect induces field-dependent rotations in the phase plane, such that an initially resonant pulse becomes off-resonant as the field builds up. To balance this effect, we detune the pulse by $\Delta=K_r\bar n_{r, s}$ such that it is resonant for the desired steady-state photon number $\bar n_{r,s}$. With this choice, the first two terms on the right of \cref{eq:EOM_Kerr_resonator} cancel each other, approximating a linear resonator in the steady state.

\begin{figure}[!t]
\includegraphics{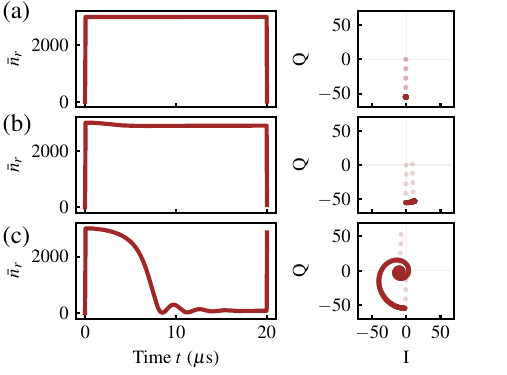}
\caption{
\label{fig:S3_NumericalSSpulse} 
Numerical simulation of the detuned three-segment pulses. (a) $\Delta=K_r\bar n_{r, s}$. (b) $\Delta=0.93 \times K_r\bar n_{r, s}$. (c) $\Delta=1.07 \times K_r\bar n_{r, s}$.
}
\end{figure}

As a result of the Kerr effect, the resonant frequency $\omega_r(n_r(t))$ changes during the ramp-up and ramp-down segments, which makes the phase and amplitude of the field deviate from those expected in the linear case. In principle, such deviations can be removed through chirped pulses where the detuning $\Delta$ is updated during the ramping, or through calibrating the phase and amplitude of the resulting state after ramping and adjusting the drive accordingly. For simplicity, we keep the same detuning and phase throughout all segments and mitigate the aforementioned problem by reducing the ramping times $t_{\uparrow}$ and $t_{\downarrow}$. From \cref{eq:rampup_amplitude}, a shorter ramping time requires a stronger amplitude, and the resulting pulse will have a broader spectrum, which makes it possible for the pulse to remain near-resonant despite the Kerr effect. We choose a \qty{40}{\ns} ramping time, which gives $1/ t_\uparrow=\qty{25}{\MHz}$. This is much larger than the frequency shift $|\omega_r(n_r=3000) - \omega_r(n_r=0)|/2\pi \approx \qty{357}{\kHz}$. Here, we take the same time length for ramp-up and ramp-down, $t_\uparrow=t_\downarrow$. We note there are also drawbacks of large amplitude ratio $\varepsilon_\uparrow/\varepsilon_s$, which may cause relatively stronger pulse distortion and also have higher requirements for the resolution of the microwave instrument, such as DACs.
% \abc{I am confused about factors of $2\pi$ here. 1/time is an angular frequency.}

We show numerical simulations of the steady-state pulses in \cref{fig:S3_NumericalSSpulse} using experimental parameters. In addition to the case where $\Delta=K\bar n_{r, s}$, we also show the results for an under-detuned pulse and an over-detuned pulse, which would be the case for possible miscalibration of the Kerr coefficient or the photon numbers. We find that the under-detuned pulse has better tolerance to such miscalibration, whereas the over-detuned pulse could easily fail to stabilize the photon numbers.

\begin{figure*}[!t]
\includegraphics{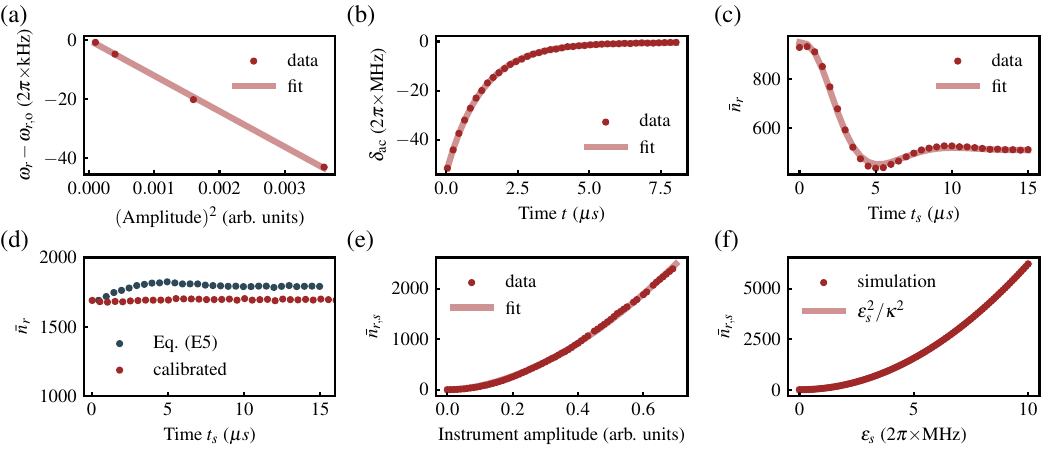}
\caption{
\label{fig: S4_Calibrations} 
Calibration results of $R_4$, which is the resonator used in \cref{fig: 4_LandauZener}. (a) Measured resonator frequency and extrapolation at different amplitudes. (b) Free decay measurement to extract the linewidth $\kappa$ of the resonator. (c) Steady-state experiment with $\Delta=0$ to extract the Kerr $K_r$ of the resonator. (d) Steady-state experiment with theoretical (blue dots) and calibrated (red dots) ramping amplitude ratio for compensating possible miscalibration and power compression of the experimental apparatus. (e) The measured and fitted relation between the instrument amplitude and the steady-state photon number $\bar n_{r,s}$. (f) The analytically calculated and numerically simulated relation between the drive amplitude $\varepsilon_s$ and the steady-state photon number $\bar n_{r,s}$.
}
\end{figure*}

%%%%%%%%%%%%%%%%%%%%%%%%%%%%%%%%%%%%%%%%
\subsection{\label{sec: LZ_pulse}Pulse in Landau-Zener sequence}

In previous sections, we explained the method to construct the pulse in our steady-state experiments. In our Landau-Zener experiments, we want to control the resonator such that the average photon number changes from $\bar n_{r, i}$ to $\bar n_{r, f}$ in a given time $t_s$. Here, we choose the ramping amplitudes such that they correspond to the ramp-up amplitude $\varepsilon_\uparrow$ in a steady-state sequence with $\bar n_{r, s}=\bar n_{r, i}$ and the ramp-down amplitude $\varepsilon_\downarrow$ in a steady-state sequence with $\bar n_{r, s}=\bar n_{r, f}$. We leave the amplitude of the quasi-steady-state segment $\varepsilon_s$ and the detuning $\Delta$ as two free parameters in a numerical optimization for the target pulses. The cost function is designed to minimize the difference between numerical results and the desired average photon numbers, which are $\bar n_{r, i}$, $\bar n_{r, f}$, and $0$ at times $t=t_{\uparrow}$, $t=t_{\uparrow}+t_s$, and $t=t_{\uparrow}+t_s+t_{\downarrow}$. For relatively short $t_s$, this method results in monotonically increasing photon number during the quasi-steady-state segment, the derivative of which can be easily extracted.

In the main text, we mention that the Landau-Zener speed is controlled by the slope of the photon number $d\bar n_r(t)/dt$. Two different parameterizations are used to adjust the slope, as each of them can only reach a limited range of Landau-Zener speeds. The time-varying parameterization, where we fix $\bar n_{r, i}=1300$ and $\bar n_{r, i}=1700$ and then change the time $t_s$, fails when $t_s$ becomes comparable to the relaxation time. As a result, the adiabatic regime cannot be reached. On the other hand, the number-varying parameterization, where we fix the time $t_s=\qty{10}{\us}$, the sum $\bar n_{r, i} + \bar n_{r, f}=3000$, and then change the difference $\bar n_{r, f} - \bar n_{r, i}$, has a maximum allowable difference $|\bar n_{r, f} - \bar n_{r, i}| < 3000$ and thus a limited diabaticity. It is the combination of the two parameterizations that enables us to explore a wide range of Landau-Zener speeds.

%%%%%%%%%%%%%%%%%%%%%%%%%%%%%%%%%%%%%%%%
\subsection{\label{sec: calibrations}Calibration procedures}

In this section, we describe the calibration procedures for the steady-state sequence and the Landau-Zener sequence. We show the results for $R_4$, the readout resonator coupled to $Q_B$, and we focus on the case where $Q_B$ is prepared in $\ket{0}$ at the beginning of the sequence.

The first step to calibrate the shaped pulse is to measure the resonator frequency at the single-photon level. We perform resonator spectroscopy for various drive amplitudes and fit each spectroscopy result to a Lorentzian function to extract a frequency~\cite{wang2025}. The single-photon frequency is then extracted by fitting these frequencies to a quadratic function of the amplitude and then extrapolating to zero amplitude, as shown in \cref{fig: S4_Calibrations}(a). The resonator spectroscopy is also used to extract the dispersive shift $\chi_{01} \equiv \omega_{r, \ket{1}} - \omega_{r, \ket{0}} = 2\pi \times \qty{-205}{\kHz}$. 

Due to the finite duration of the spectroscopy pulse, the fitted quality factor may have systematic errors. In such cases, a free decay experiment is preferable to extract the linewidth $\kappa$ of the resonator. At any time $t$, we use transmon spectroscopy to measure the ac-Stark shift $\delta_{\rm ac}(t)$ of the $\ket{0}\leftrightarrow\ket{1}$ transition frequency of the coupled transmon. These ac-Stark shifts are then fitted to an exponential function, as shown in \cref{fig: S4_Calibrations}(b), which gives $\kappa_{\ket{0}} = 2\pi \times \qty{127}{\kHz}$.

Using the fitted linewidth, we can calculate the ramping amplitude ratio in \cref{eq:rampup_amplitude}. The Kerr coefficient is then measured by applying the three-segment pulse. Here, we set $\Delta=0$ without any prior knowledge of the value of the Kerr coefficient, and the photon number $\bar n_r$ is not stabilized. In that case, the Kerr effect manifests through the time-dependence of $\bar n_r(t)$. We calculate the measured photon number using
\begin{equation}
\bar n_r(t)=\delta_{\rm{ac}}(t)/\chi_{01},
\label{eq:photons_stark_shift}
\end{equation}
The results are fitted to the numerical simulation of the three-segment pulse, where the Kerr value $K_r$, the effective drive amplitude $\varepsilon_s$, and the detuning $\Delta$ are treated as fitting parameters, as shown in \cref{fig: S4_Calibrations}(c). We leave the detuning $\Delta$ as a free parameter to further correct the single-photon frequency extrapolated from \cref{fig: S4_Calibrations}(a). As a result, we find $K_{r, \ket{0}}=2\pi \times \qty{-119}{\Hz}$.

After calibrating $\omega_r$, $\kappa$, $K_r$, and $\varepsilon_s$, we have the minimal parameters to run the steady-state experiment: we can set $\Delta=K_r(\varepsilon_s/\kappa)^2$ and choose the ramping amplitude ratio based on \cref{eq:rampup_amplitude}. The result is shown as blue dots in \cref{fig: S4_Calibrations}(d). Although the photon number is stable for a long time, it does not reach its steady state immediately after the ramp-up segment because the actual amplitude ratio required to be applied on the device may deviate from the theoretical value $\varepsilon_\uparrow/\varepsilon_s \approx 63.3$ that we set to the instrument. Such deviation may come from miscalibration of the linewidth $\kappa$ or the power compression of the instrument. We correct it by empirically adjusting the amplitude ratio to $\varepsilon_\uparrow/\varepsilon_s = 68.5$, and the result is shown as red dots in \cref{fig: S4_Calibrations}(d).

Another consequence of power compression is that the effective drive amplitude $\varepsilon_s$ is not proportional to the instrument amplitude. We perform the steady-state experiment at different instrument amplitudes and extract the corresponding steady-state photon numbers $\bar n_{r,s}$. The results are fitted to a phenomenological model $y(x)=Ax^m$, and we get $m \approx 1.776 \neq 2$, as shown in \cref{fig: S4_Calibrations}(e). This relation gives us the conversion between the instrument amplitude and the $\bar n_{r,s}$ reported in \cref{fig: 4_LandauZener}(b). We also calculate a similar conversion between the steady-state photon number $\bar n_{r,s}$ and the drive amplitude $\varepsilon_s$ on the resonator using numerical simulation, as shown in \cref{fig: S4_Calibrations}(f). The results agree well with analytical prediction $\bar n_{r,s} = \varepsilon_s^2 / \kappa^2$.

%%%%%%%%%%%%%%%%%%%%%%%%%%%%%%%%%%%%%%%%
\subsection{\label{sec:photon_numbers}Errors in calibration of photon numbers}

In the calibration procedures discussed in \cref{sec: calibrations}, the photon numbers are extracted using \cref{eq:photons_stark_shift}. This relation is based on the dispersive Hamiltonian. For a multilevel system coupled to a single-mode harmonic oscillator, the dispersive Hamiltonian up to sixth order in perturbation is ($\hbar=1$)
\begin{equation}
\begin{split}
\hat{H}_{\rm{disp}} &= \sum_j \omega_j \ket{j}\bra{j} + \omega_r \hat{a}^{\dagger}\hat{a} \\
&+ \sum_j \chi_j \hat{a}^{\dagger}\hat{a} \ket{j}\bra{j} 
+ \sum_j \frac{\eta_j}{2} \hat{a}^{\dagger}\hat{a}^{\dagger}\hat{a}\hat{a} \ket{j}\bra{j}\\
&+ \sum_j \frac{\mu_j}{6} \hat{a}^{\dagger}\hat{a}^{\dagger}\hat{a}^{\dagger}\hat{a}\hat{a}\hat{a} \ket{j}\bra{j}.
\label{eq:full_dispersive_hamiltonian}
\end{split}
\end{equation}
Below, we discuss possible errors when using this equation.

%%%%%%%%%%%%%%%%%%%%%%%%%%%%%%%%%%%%%%%%
\subsubsection{\label{sec:Kerr_effect}Kerr effect}

The dispersive Hamiltonian in \cref{eq:full_dispersive_hamiltonian} is often truncated to the second order, with the dispersive shift defined as $\chi_{01}=\chi_1-\chi_0$. However, at large photon numbers, the state-dependent four-wave mixing Kerr $\eta_j$ and six-wave mixing Kerr $\mu_j$ also have non-negligible contributions to the dispersive shift. In other words, the denominator of \cref{eq:photons_stark_shift} depends on photon number, and the relation between photon number and ac-Stark shift is therefore not linear at large photon numbers.

\begin{figure}[!b]
\includegraphics{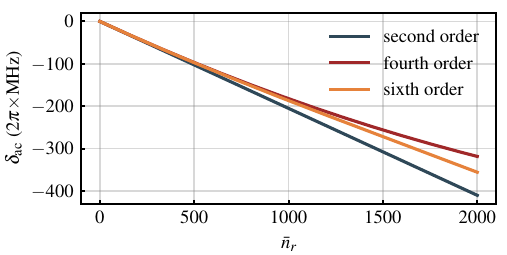}
\caption{
\label{fig: S5_6WaveMixingKerr} 
Estimated ac-Stark shift as a function of photon number for different orders in perturbation theory.
}
\end{figure}

To estimate the possible errors from these higher-order effects, we calculate the ac-Stark shift $\delta_{\rm{ac}}$ using $\eta_j$ and $\mu_j$ extracted from numerical diagonalization with the parameters shown in \cref{tab: device_params}. The results are shown in \cref{fig: S5_6WaveMixingKerr}. We find that at 1500 photons, the critical photon number in Fig. 4 of the main text, there could be an underestimation of the photon number of about 100 to 200 photons for a given $\delta_{\rm{ac}}$.

Because of the existence of $\mu_j$ and other higher-order Kerr effects, our choice of detuning for the steady-state experiments, $\Delta=K_r\bar n_{r, s}$ as explained in \cref{sec: Kerr_resonator}, fails to stabilize the resonator state at higher photon numbers. An ideal choice of detuning should take into account the nonlinear relation between the dispersive shift and the photon number.

%%%%%%%%%%%%%%%%%%%%%%%%%%%%%%%%%%%%%%%%
\subsubsection{\label{sec:averaged_photon_numbers}Averaged photon numbers}

Suppose that we truncate \cref{eq:full_dispersive_hamiltonian} to second order and treat the transmon as a two-level system. In the rotating frame, the Hamiltonian of the system under the spectroscopy pulse can then be simplified to
\begin{equation}
\hat{H}_{\rm{spec}}(t) 
= \frac{1}{2} \left[ \Delta-\chi_{01}n_r(t)\right] \sigma_z
+ \frac{1}{2}\Omega(t)\sigma_x.
\label{eq:spectroscopy_hamiltonian}
\end{equation}
Here, $\Delta \equiv \omega_d - \omega_{01}$ is the detuning of the spectroscopy pulse to the qubit frequency, and $\Omega(t)$ is the pulse envelope. In the above expression, we have performed the rotating-wave approximation (RWA). The ac-Stark shift $\delta_{\rm{ac}}(t)$ is measured from the excited population $P_{\ket{1}}$ after a spectroscopy pulse of duration $T$. As a result, the measured $\delta_{\rm{ac}}(t)$ reflects the averaged photon number over the time window of the spectroscopy pulse. To see this, we perturbatively calculate the dynamics of the system under \cref{eq:spectroscopy_hamiltonian} using average Hamiltonian theory~\cite{brinkmann2016}. In this framework, the propagator $U$ is given by
\begin{equation}
\begin{split}
U(T, 0) &= e^{-i\bar{H}_{\rm{spec}}T}, \\
\bar{H}_{\rm{spec}} &= \bar{H}_{\rm{spec}}^{(1)} + \bar{H}_{\rm{spec}}^{(2)} + ...,
\label{eq:AHT}
\end{split}
\end{equation}
where the time-averaged Hamiltonian $\bar{H}_{\rm{spec}}$ is expanded at each order in the drive amplitude $\Omega(t)$. Suppose that the system is prepared in the ground state $\ket{0}$ before the spectroscopy. The excitation probability to lowest-order in $\Omega(t)$ is
\begin{equation}
\begin{split}
P_{\ket{1}} &= |\bra{1}e^{-i\bar{H}_{\rm{spec}}^{(1)}T}\ket{0}|^2 \\
&= 4 \theta^2 {\rm sinc}^2 \left( \frac{1}{2}\sqrt{\theta^2 + T^2 (\Delta-\chi_{01}\bar n_r)^2} \right).
\label{eq:AHT_P1}
\end{split}
\end{equation}
Here, we introduced the rotation angle $\theta \equiv \int_0^T \Omega(t) dt$ and the time-averaged photon number $\bar n_r \equiv 1/T \int_0^T n(t)dt$. The spectroscopic response peaks at $\Delta=\chi_{01}\bar n_r$, which is proportional to the time-averaged photon number instead of the instantaneous photon number. Hence, when we fit the Kerr value $K_r$ in \cref{fig: S4_Calibrations}(c), the numerical simulation results are uniformly averaged over a time window with duration $T$. Going to higher orders in perturbation, however, the excitation probability $P_{\ket{1}}$ depends on the specific shape of $\Omega(t)$. This leads to errors in the fitted Kerr value.

We also note that the resonator state generated by a classical drive is usually not a Fock state. The photon number discussed here should thus be thought of as a weighted average over different Fock states. In the number-splitting regime where the coupling between the transmon and the resonator is strong, $\chi_{01}>1/T$, there is more than one peak in the spectroscopic response~\cite{schuster2007}.

%%%%%%%%%%%%%%%%%%%%%%%%%%%%%%%%%%%%%%%%
\section{\label{sec: full_fig4b}Identification of final states in the steady-state experiment}

In \cref{sec: LandauZener}, we mentioned that $\ket{6}$ is one of the final states in the steady-state experiment. Here we give more details on this identification.

\begin{figure}[!ht]
\includegraphics{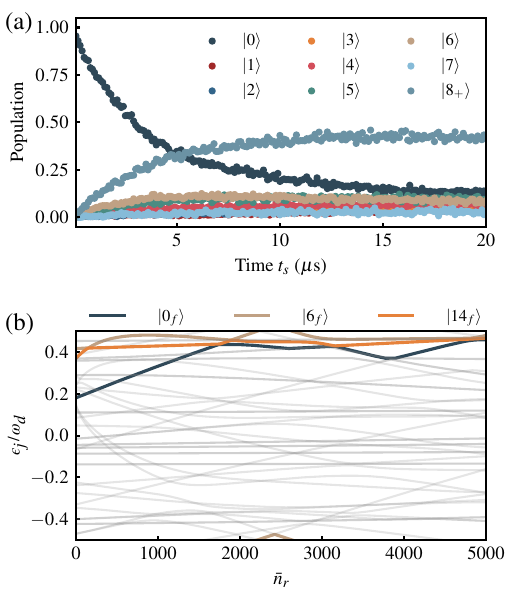}
\caption{
\label{fig: S6_full_fig4b}
Identification of final states in the steady-state experiment. (a) A linecut of the steady-state experiment results shown in \cref{fig: 4_LandauZener}(b) at 1500 photons. (b) Normalized Floquet quasienergies $\epsilon_j/\omega_d$ for $Q_B$. We take $n_g=0$.
}
\end{figure}

As shown in \cref{fig: S6_full_fig4b}(a), we \bd{find} apparent population increase in $\ket{6}$, leaving $\ket{7}$ nearly unaffected. The large population observed in states $\ket{8}$ or higher suggests that the population in state $\ket{6}$ is then further transferred to higher excited states. These observations are consistent with the Floquet analysis shown in \cref{fig: S6_full_fig4b}(b). Following the branch $\ket{0_f}$, we encounter two nearby avoided crossings around $\bar n_r \sim 1800$, involving both branches $\ket{6_f}$ and $\ket{14_f}$. During the long steady-state segment, the population can be transferred to either of these branches. The $\ket{6_f}$ and $\ket{14_f}$ branches also show an avoided crossing at much lower photon number ($\bar n_r \sim 100$). As a result, in the experiment, the diabatic ramp-down segment leads to the population in branch $\ket{14_f}$ being assigned as $\ket{6}$ and to the population in branch $\ket{6_f}$ being assigned as $\ket{8_+}$.

Notice that the critical photon number shown in \cref{fig: S6_full_fig4b} is slightly larger than the measured results ($\sim 1500$ photons) in \cref{fig: 4_LandauZener}. This is likely due to the effect of higher-order nonlinear terms that are accounted for in the Floquet branch analysis but not in our experimental calibration \rr{of the steady-state sequence}; see \cref{sec:Kerr_effect}. \rr{All photon numbers reported in \cref{sec: typical_transmons} have been corrected for this effect.}

\rr{
%%%%%%%%%%%%%%%%%%%%%%%%%%%%%%%%%%%%%%%%
\section{\label{sec: FBA}Floquet branch analysis}
}

The Floquet branch analysis for studying transmon ionization has been previously explained in detail in Ref.~\cite{dumas2024}. Here, we add a few supplements related to this work.

%%%%%%%%%%%%%%%%%%%%%%%%%%%%%%%%%%%%%%%%
\subsection{\label{sec: LZ_speed}Landau-Zener speed}

For an avoided crossing at $n_{r, \rm crit}$, \bd{the photon-number slope} $dn_r(t)/dt|_{n_{r, \rm crit}}$ can be extracted from experiments or from the numerical solution of \cref{eq:EOM_Kerr_resonator}. \bd{Moreover,} the Floquet \bd{quasienergies $\epsilon_j(n_r)$ can be} calculated as a function of the photon number $n_r$. The Landau-Zener speed $v$ \bd{is then approximately given by}
\begin{equation}
    v = \sqrt{2\Delta_{\rm ac} \frac{d^2\epsilon_j(n_r)}{dn_r^2} \Big|_{n_{r, \rm crit}}} \frac{dn_r(t)}{dt}\Big|_{n_{r, \rm crit}},
\end{equation}
\bd{where $\Delta_{\rm ac}$ is the quasienergy gap at the avoided crossing}. If the Floquet \bd{quasienergies are instead} calculated as a function of effective transmon drive amplitudes $\varepsilon_t \equiv 2g\sqrt{n_r}$, the Landau-Zener speed becomes
\begin{equation}
    v = \sqrt{\frac{2g^2\Delta_{\rm ac}}{n_{r, \rm crit}} \frac{d^2\epsilon_j(\varepsilon_t)}{d\varepsilon_t^2} \Big|_{\varepsilon_{t, \rm ac}}} \frac{dn_r(t)}{dt} \Big|_{n_{r, \rm crit}}.
\end{equation}
We also note that \cref{eq: FLZ} is derived assuming $\hbar=1$, so both $\Delta_{\rm ac}$ and $v$ should have units of angular frequency when calculating the transition probability.

%%%%%%%%%%%%%%%%%%%%%%%%%%%%%%%%%%%%%%%%
\subsection{\label{sec: Floquet_frequencies}Frequency dependence of the Floquet spectrum}

The simulated $n_{r,\rm crit}$ values shown in \cref{fig: 5_timetrace}(d) and \cref{fig: 6_ng}(a) are extracted from Floquet spectra calculated over different $n_g$ values, assuming a fixed drive frequency. In contrast, the steady-state experiment in \cref{fig: 5_timetrace}(b), used to measure $n_{r,\rm crit}$ is performed with varying drive detuning $\Delta=K_r \bar n_{r,s}$ for each $\bar n_{r,s}$. We justify the validity of this comparison by noting that a small change in the drive frequency, $\delta \omega_d$ does not significantly alter the main features of the Floquet spectrum as long as $\delta \omega_d \ll \omega_d, \omega_{j,j+1}$

% The steady-state experiment used to measure $n_{r,\rm crit}$ in \cref{fig: 5_timetrace}(b) is performed with a different drive detuning $\Delta=K_r \bar n_{r,s}$ for each $\bar n_{r,s}$. For simplicity, however, the simulated $n_{r,\rm crit}$ in \cref{fig: 5_timetrace}(d) and \cref{fig: 6_ng}(a) are extracted from Floquet spectra \bd{by varying} $n_g$ \bd{but keeping the drive frequency constant}. \bd{The assumption of a constant drive frequency does not change the main features of the Floquet spectrum as long as small changes $\delta \omega_d \sim \Delta$ of the drive frequency remain smaller than the drive frequency and transmon transition frequencies, $\delta \omega_d \ll \omega_d, \omega_{j,j+1}$.}
% We justify the effectiveness of this comparison by noting that a small change of the drive frequency $\delta \omega_d$ does not significantly affect the major features in the Floquet spectrum, when such a change is much smaller than the drive frequency and transmon transition frequencies, $\delta \omega_d \ll \omega_d, \omega_{j,j+1}$.

\begin{figure}[ht]
\includegraphics{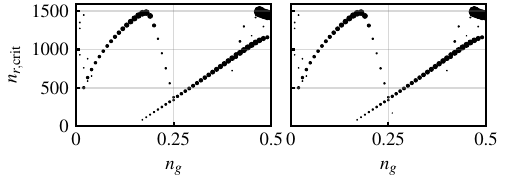}
\caption{
\label{fig: S9_Floquet_frequencies}
Offset charge dependence of the critical photon number. We show simulation results for two different drive frequencies, corresponding to the stimulation frequencies in the steady-state experiments of $Q_c$ with $\bar n_{r,s} \approx 1180$ (left) and $\bar n_{r,s} \approx 680$ (right). The difference of the two drive frequencies is about $\delta \omega_d / 2\pi \approx \qty{28}{\kHz}$. The marker sizes are the same as in \cref{fig: 5_timetrace}(d). The two simulations agree with each other except for the smallest gap at $n_g=0.26$}
\end{figure}

In \cref{fig: S9_Floquet_frequencies}, we show the extracted $n_{r,\rm crit}$ as a function of $n_g$ for two drive frequencies separated by $\delta \omega_d / 2\pi \approx \qty{28}{\kHz}$. Overall, they agree with each other except for the smallest gap at $n_g=0.26$ with $\Delta_{\rm ac}/2\pi \approx \qty{11}{\kHz}$.

%%%%%%%%%%%%%%%%%%%%%%%%%%%%%%%%%%%%%%%%
\subsection{\label{sec: Floquet_increments}Choice of increment when calculating the Floquet spectrum}

When two Floquet branches cross each other in a Floquet spectrum, they form an avoided crossing as long as there is a nonzero coupling between them. In numerical simulations, an avoided crossing with a small gap can only be identified when the resolution of the photon number (or of any other sweeping parameters) is sufficiently high, i.e., \bd{we choose a} small \bd{enough} $\delta n_r$. Otherwise, the two branches will not form an avoided crossing.

\begin{figure}[ht]
\includegraphics{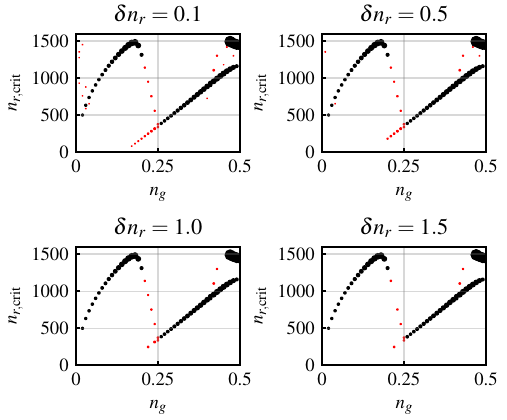}
\caption{
\label{fig: S10_Floquet_increments}
The critical photon numbers for different increments $\delta n_r$ when performing Floquet analysis as in the \cref{fig: 6_ng}(a). Avoided crossings with $\Delta_{\rm ac}/2\pi < \qty{1}{\MHz}$ are shown in red to highlight the differences between the subfigures.
}
\end{figure}

We show this numerical effect in \cref{fig: S10_Floquet_increments}. Here, for each $n_g$, the Floquet spectrum is calculated for photon numbers from 0 to 1600 with increment $\delta n_r$. Avoided crossings with $\Delta_{\rm ac}/2\pi < \qty{1}{\MHz}$ are shown in red to highlight the differences between the subfigures. As expected, increasing this increment \bd{misses} small gaps in the spectrum. When doing such a simulation, a proper increment can then be chosen based on the relevant gap size involved in the experiment and analysis. We note that the proper value of $\delta n_r$ may strongly depend on the transmon parameters and the drive frequency, and our results here should be viewed only as an \bd{illustrative} example \bd{for transmon} $Q_c$.

%%%%%%%%%%%%%%%%%%%%%%%%%%%%%%%%%%%%%%%%
\subsection{\label{sec: Floquet_Q4Q0}Comparing offset charge dependency between a typical transmon and a high-$E_J/E_C$ transmon}

The experiment shown in \cref{fig: 4_LandauZener}(b) took a few hours, but the measured critical photon number of $Q_B$ was fairly stable and reproducible, and we didn't observe strong fluctuations as for the results shown in \cref{fig: 5_timetrace}(b). These distinct behaviors come from the \bd{very different $E_J/E_C$ ratios}
%and the spectra
of $Q_B$ and $Q_C$.

\begin{figure}[ht]
\includegraphics{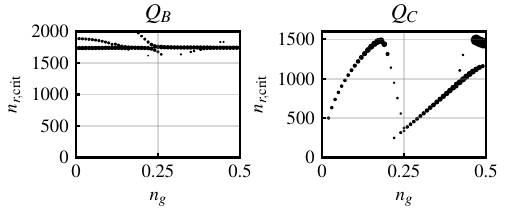}
\caption{
\label{fig: S11_Floquet_Q4Q0}
Offset charge dependency of the critical photon numbers for a high-$E_J/E_C$ transmon $Q_B$ (left) and a typical transmon $Q_C$ (right). In these Floquet analyses, we use $\delta n_r=1.0$.
}
\end{figure}

For a high $E_J/E_C$ transmon, for example $Q_B$, the final state $\ket{6}$ is insensitive to the charge noise. As a result, the corresponding Floquet mode $\ket{6_f}$ only weakly depends on $n_g$. In its Floquet spectrum, the relatively stable structure of branches $\ket{0_f}$ and $\ket{6_f}$ ensures that the position $n_{r,\rm crit}$ and gap size $\Delta_{\rm ac}$ of some avoided crossings remain the same over \bd{a wide range of} $n_g$, as shown in the left panel of \cref{fig: S11_Floquet_Q4Q0}. This is not true for a typical transmon like $Q_C$, \bd{whose lower $E_J/E_C$ leads} $n_{r,\rm crit}$ and $\Delta_{\rm ac}$ \bd{to both} strongly depend on $n_g$.

\clearpage

% %%%%%%%%%%%%%%%%%%%%%%%%%%%%%%%%%%%%%%%%
% \section{\label{sec: ng_conversion}Relation between $n_g$ and $\delta \omega_{12}$}

% \begin{figure}[!ht]
% \includegraphics{figS12.pdf}
% \caption{
% \label{fig: S12_ng_conversion}
% Frequency difference $\delta \omega_{12}$ between the even and odd charge parity as a function of $n_g$ for the $\ket{1} \leftrightarrow \ket{2}$ transition of $Q_C$.
% }
% \end{figure}

%%%%%%%%%%%%%%%%%%%%%%%%%%%%%%%%%%%%%%%%
\bibliography{Ionization}

\end{document}